\documentclass[conference,10.5pt]{IEEEtran}
\IEEEoverridecommandlockouts
\usepackage{cite}
\usepackage{amsmath,amssymb,amsfonts}
\usepackage{algorithm}
\usepackage{algcompatible}

\algnewcommand\algorithmicreturn{\textbf{return}}
\algnewcommand\RETURN{\State \algorithmicreturn}%
\usepackage{graphicx}
\usepackage{color}
\usepackage{subfig}
\usepackage{stackengine}
\usepackage{comment}
\usepackage[normalem]{ulem}
\usepackage{tabularx}
\usepackage{bm}
\usepackage{bbold}
\usepackage{mathtools, nccmath}
\usepackage{multirow}
\usepackage{stfloats}

\usepackage[singlespacing]{setspace} 

\usepackage{lettrine}
\usepackage[font=scriptsize]{caption}

\usepackage[colorlinks, citecolor=blue]{hyperref} 

\usepackage[flushleft]{threeparttable} 
\usepackage{booktabs}

\usepackage{textcomp}
\usepackage{xcolor}

\newtheorem {theorem}{Theorem}

\def\BibTeX{{\rm B\kern-.05em{\sc i\kern-.025em b}\kern-.08em
    T\kern-.1667em\lower.7ex\hbox{E}\kern-.125emX}}
    
\DeclareMathOperator*{\argmax}{argmax}

\usepackage{orcidlink}

\usepackage{xspace}
\usepackage{ifthen}
\usepackage[inline]{enumitem}

\newboolean{showcomments}
\setboolean{showcomments}{true}
\ifthenelse{\boolean{showcomments}}
{
}

\usepackage{subfig}
\usepackage{fancyhdr}

\newcommand*{\affmark}[1][*]{\textsuperscript{#1}}

\begin{document}

\title{\huge Explainable Multi-Agent Reinforcement Learning for Extended Reality Codec Adaptation}

\author{\small\IEEEauthorblockN{Pedro Enrique Iturria-Rivera\orcidlink{0000-0002-8757-2639}\affmark[1], \IEEEmembership{Student Member,~IEEE}, Raimundas Gaigalas\affmark[2],  Medhat Elsayed\orcidlink{0000-0002-1106-6078}\affmark[2], \\ \IEEEmembership{Senior Member,~IEEE}, Majid Bavand\affmark[2], Yigit Ozcan\affmark[2] and Melike Erol-Kantarci\orcidlink{0000-0001-6787-8457}\affmark[1], \IEEEmembership{Senior Member,~IEEE}}
\IEEEauthorblockA{\affmark[1]\textit{School of Electrical Engineering and Computer Science, University of Ottawa, Ottawa, Canada}\\\affmark[2]\textit{Ericsson Inc., Ottawa, Canada}}Emails:\{pitur008, melike.erolkantarci\}@uottawa.ca, \{raimundas.gaigalas, medhat.elsayed,\\ majid.bavand, yigit.ozcan\}@ericsson.com\vspace{-1em} }

\maketitle
\begin{abstract}
\textbf{
Extended Reality (XR) services are set to transform applications over $5^{th}$ and $6^{th}$ generation wireless networks, delivering immersive experiences. Concurrently, Artificial Intelligence (AI) advancements have expanded their role in wireless networks, however, trust and transparency in AI remain to be strengthened. Thus, providing explanations for AI-enabled systems can enhance trust. We introduce Value Function Factorization (VFF)-based  Explainable (X) Multi-Agent Reinforcement Learning (MARL) algorithms, explaining reward design in XR codec adaptation through reward decomposition. We contribute four enhancements to XMARL algorithms. Firstly, we detail architectural modifications to enable reward decomposition in VFF-based MARL algorithms: Value Decomposition Networks (VDN), Mixture of Q-Values (QMIX), and Q-Transformation (Q-TRAN). Secondly, inspired by multi-task learning, we reduce the overhead of vanilla XMARL algorithms. Thirdly, we propose a new explainability metric, Reward Difference Fluctuation Explanation (RDFX), suitable for problems with adjustable parameters. Lastly, we propose adaptive XMARL, leveraging network gradients and reward decomposition for improved action selection. Simulation results indicate that, in XR codec adaptation, the Packet Delivery Ratio reward is the primary contributor to optimal performance compared to the initial composite reward, which included delay and Data Rate Ratio components. Modifications to VFF-based XMARL algorithms, incorporating multi-headed structures and adaptive loss functions, enable the best-performing algorithm, Multi-Headed Adaptive (MHA)-QMIX, to achieve significant average gains over the Adjust Packet Size baseline up to $10.7\%$, $41.4\%$, $33.3\%$, and $67.9\%$ in XR index, jitter, delay, and Packet Loss Ratio (PLR), respectively.
}

\end{abstract}


\section{Introduction}

Applications of Artificial Intelligence (AI) and Machine Learning (ML) across many industries have been on the rise recently and the new generations of wireless networks $5^{th}$ (5G) and $6^{th}$ (6G) have not been exceptions. AI presents great potential in optimizing numerous functionalities in wireless networks and can provide autonomy in situations where human intervention is not feasible due to high complexity. This capability has led to the integration of AI-based algorithms in many critical applications within the telecommunications industry. These algorithms are often essential components of network architecture, including deep neural networks, which are frequently considered black boxes from a human perspective \cite{adadi2018peeking}. Evidently, this raises concerns regarding trust. Additionally, faulty human design can significantly impact confidence in AI-based systems, especially during the early design stages. The necessity for trust in AI systems has been a recurring topic among researchers, and due to its extensive use, it has had an impact on society. For instance, in June 2022, the Canadian Government presented the AI and Data Act (AIDA) \cite{TheArtif85:online} where it is stated that ``the design, development, and use of AI systems must be safe, and must respect the values of Canadians''. Similar statements have been proposed around the world, leading the research community to seek ways to guarantee trust and transparency in the usage of AI. For this reason, providing explanations or interpretations of current AI-based solutions using Explainable AI (XAI) is called for. To respond to this call several proposals have been made to find explanations for existing algorithms in the subfield of machine learning. Some comprehensive studies in the literature attempt to classify these efforts, with supervised machine learning algorithms being the most commonly studied area at the moment of writing this paper \cite{burkart2021survey}. However, a handful of surveys exist in the emerging subfield of explainable reinforcement learning  \cite{vouros2022explainable, hickling2023explainability, milani2024explainable}. A common categorization divides explainable machine learning methods into two categories: intrinsic and post-hoc \cite{hickling2023explainability}. Intrinsic interpretability involves inherently understandable building models, like small decision trees. In contrast, post-hoc interpretability involves developing a supplementary model to explain an existing one. Despite the valuable information provided by post-hoc methods, all explanations are acquired after obtaining a trained model, making it difficult to gather information online. Additionally, the explainability of Multi-Agent Reinforcement Learning (MARL) algorithms is an even more under-researched area \cite{boggess2023explainable}, with only a few papers existing in the literature due to its increased complexity compared to single-agent approaches.

6G can be identified by two main features compared to its predecessors: sustainability and proactiveness \cite{khan2022digital}. This means that 6G must be equipped with self-sustainability and adapt to the newly offered services bounded to high requirements and user demand. Extended Reality (XR) and Cloud Gaming (CG) are considered among the services that 6G will support. XR has indeed been previously discussed in 3GPP 5G-related documentation \cite{3gpp1, 3gpp2}, however, the leap of 6G in terms of the usage of AI/ML \cite{bang20236g} will impact immersive technologies\cite{Fiverese47:online} as well. 

In this work, we intend to study XR traffic codec adaptation using Value Function Factorization (VFF)-based Explainable (X) MARL algorithms. To comply with recent regulations to improve trust in AI-based systems, we delve into Explainable MARL (XMARL) and extend  Reward Decomposition \cite{juozapaitis2019explainable} in state-of-the-art VFF-based MARL algorithms. Furthermore, network latency, packet loss, and jitter are considered the main network-related Key Performance Indicators for XR services \cite{itug1035}. The implications of the previous metrics in the user experience have been studied consistently. For such a reason in this work, the performance results of the proposed algorithms are shown utilizing the mentioned KPIs.  This work differs from the previous works with the following four contributions:
\begin{enumerate}

    \item We present a general architecture to enable reward decomposition in MARL algorithms. 
    \item We propose alternative solutions to reduce the overhead of vanilla decomposed MARL  algorithms using multi-headed structures.
    \item We introduce a new metric called Reward Difference Fluctuation Explanations (RDFX) fitted for problems where actions correspond to parameters to optimize.
    \item Finally, we present an online/adaptive/explainable solution that uses network gradients and leverage reward decomposition to improve the action selection of the XMARL algorithms. 
\end{enumerate}

These contributions enable us to draw conclusions regarding the design of composite rewards in VFF-based MARL algorithms. In this paper, we focus on explaining XR codec adaptation. However, our findings are agnostic to the use case and can be extended to other applications. By decomposing the objective function, and consequently, the Q-Value functions, we can observe the behavior of each component. This observation led us to propose an adaptive approach that does not require post-hoc analysis to determine the importance of rewards. Specifically, simulation results demonstrate that the best-performing algorithm, Multi-Headed Adaptive (MHA)-QMIX, shows improvement in all studied network Key Performance Indicators (KPIs) over the Adjust Packet Size (APS) algorithm, a baseline introduced in \cite{Bojovic2023, Lagen2023}, by up to $10.7\%$, $41.4\%$, $33.3\%$, and $67.9\%$ in XR index, jitter, delay, and Packet Loss Rate (PLR) respectively.

The rest of this paper is organized as follows. Section \ref{Section2} presents the existing works related to XR traffic codec adaptation, video streaming optimization, and explainable AI in the wireless context.  Section \ref{section3} introduces some background on MARL, Explainable Reinforcement Learning,  VFF, and  Quality of Experience (QoE) in XR in next-generation networks. In Section \ref{section4} a brief system model is described. Additionally, section \ref{section5} presents the simulation results of our proposed scheme as well as the presented baseline. Additionally, we present a list of potential applications in the telecommunications industry. Finally, Section \ref{Section6} concludes the paper.

\section{Related work}\label{Section2}
In this work, we study two areas of research: explainability in MARL and its application in XR codec adaptation. Recently, explaining or interpreting RL in the context of wireless networks has become of interest, marking it as an emerging area. However, to the best of our knowledge, there has been no studies in the area of reward decomposition in wireless networks, in which we focus, by tackling MARL explainability for XR codec adaptation. Next, we present a list of related studies to our research below. 

Recently in \cite{iturria}, the authors proposed a MARL algorithm to perform cross-optimization of XR codec parameters. Such a proposal consists of a cooperative Multi-Agent System (MAS) based on an Optimistic Mixture of Q-Values (oQMIX). It leverages an attention mechanism and slate-Markov Decision Process (MDP) to improve the oQMIX algorithm’s action selection. The previous work uses a baseline called Adjust Packet Size (APS) introduced in \cite{Bojovic2023, Lagen2023} where the authors focus on the exploration of innovative Quality of Service (QoS) management strategies to address the unique characteristics of XR traffic. 

In terms of explainability, a few efforts have been made to offer explanations for AI/ML in the context of wireless networks. Recently, \cite{fiandrino2023explora} proposed a framework to provide explanations for Deep Reinforcement Learning (DRL)-based resource allocation solutions in cellular networks. It offers post-hoc explanations using a graph that establishes a link between the actions taken by a DRL agent and the input state space. Although this framework was initially proposed in the context of Open Radio Access Network (O-RAN), the authors claim it can effectively generalize to other applications. In \cite{he2021explainable}, the authors propose a deep neural network-based path planner for Unmanned Aerial Vehicles (UAVs). Additionally, they introduce a model explanation method to provide post-hoc textual and visual explanations for the behavior of UAVs. Recent surveys \cite{guo2020explainable, fiandrino2022toward, brik2023survey} have delved into the importance of reasoning and explaining the usage of AI/ML in new-generation wireless networks. On one side, \cite{guo2020explainable} highlights the need to employ XAI in 6G networks to ensure trust in ML methods. In \cite{fiandrino2022toward}, the authors build upon previous work and present a list of tools to explain wireless networks, along with the current challenges of XAI in 6G. Finally, in \cite{brik2023survey}, the authors provide an overview of the current state of the art in XAI and the potential applications of such techniques in the novel O-RAN. In the next section, we present the technical background for our proposed algorithms.

\section{Background and System Model} \label{section3}
In this section, we present background information on Reward Decomposition (RD) and VFF in MARL including the theoretical basis for our XMARL algorithms. Additionally, we introduce some aspects of QoE in XR which motivates our research as an application area. 

\subsection{Explainable Multi-Agent Reinforcement Learning Algorithm Taxonomy}
\begin{figure}[t]
\center
  \includegraphics[scale=0.70]{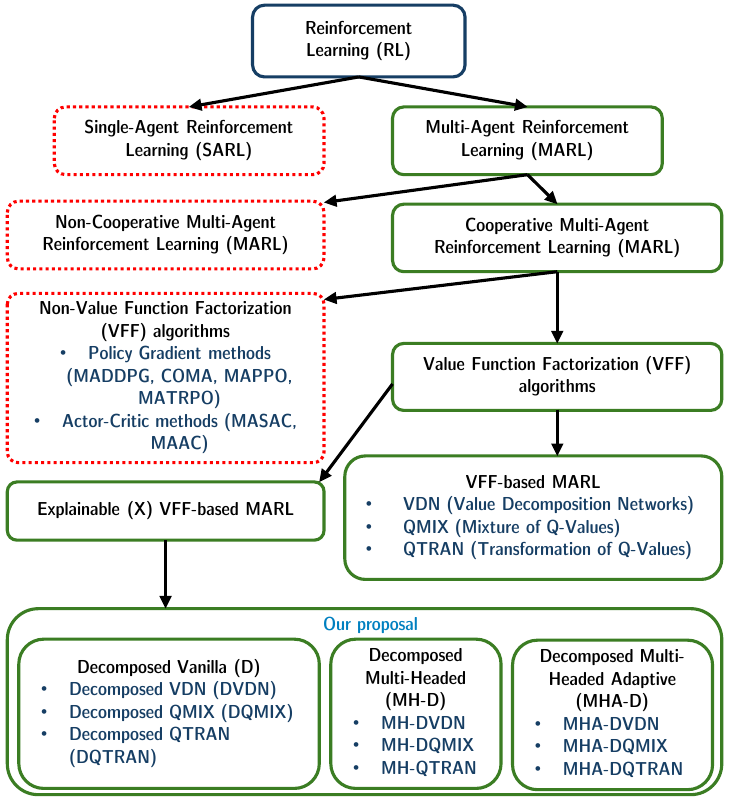}
  \setlength{\belowcaptionskip}{-5pt}
  \caption{ High-level taxonomy of proposed algorithms. The red colored-dashed boxes depict the area of reinforcement learning out of the scope of this work. At the bottom of the figure, the three explainable MARL families that are proposed as part of this paper.} 
  \label{taxonomy}

\end{figure}

In this subsection, we introduce a high-level taxonomy of our proposed algorithms. Figure \ref{taxonomy} depicts the high-level categories of the taxonomy and their relationship to the RL framework. The red dashed boxes represent areas of reinforcement learning that are out of the scope of this paper. At the root of the tree, MARL algorithms are split into two main categories: cooperative and non-cooperative. Non-cooperative includes independent and competitive approaches. Meanwhile, cooperative MARL includes VFF methods and non-VFF methods, such as Policy Gradient-based algorithms like MADDPG \cite{lowe2017multi}, COMA \cite{foerster2018counterfactual}, MAPPO \cite{yu2022surprising}, and MATRPO \cite{li2023multiagent}, as well as Actor-Critic methods like MASAC \cite{pu2021decomposed} and MAAC \cite{iqbal2019actor}. Note that additional algorithms can be classified within these categories, but we have only mentioned the most cited. Finally, at the bottom of the tree, we present our proposed algorithms, inspired by VFF-based MARL algorithms. Three families are depicted, each corresponding to the main algorithms proposed in this paper: Decomposed Vanilla (D), Decomposed Multi-Headed (MH-D), and Multi-Headed Adaptive (MHA-D) algorithms, respectively. More details on each family is presented in the following sections. 

\subsection{Markov Decision Model for Cooperative Multi-Agent Tasks}

Let's define a fully cooperative multi-agent task as a Decentralized Partially Observable Markov Decision Process (Dec-POMDP) \cite{oliehoek2016concise} composed of the tuple $\mathcal{G} = <\mathcal{S}, \mathcal{A}, P, r, \mathcal{Z}, O, N, \gamma>$ where $\bm{s} \in \mathcal{S}$ corresponds to the state of the environment. In this setup, each agent $n \in \mathcal{N} : = \{1, ..., N\}$ selects an action $a_n \in \mathcal{A}$ at each time step. After all agents' action decisions, a joint action vector $\bm{a} \in \mathcal{A}$ is obtained. The probability transition function can be defined as $P(\bm{s}', \bm{s}, \bm{a}): \mathcal{S} \times \mathcal{A}^N \times \mathcal{S} \rightarrow [0,1]$ and $\gamma \in [0,1]$ is a discount factor to give less importance to future rewards. In this context, all agents share their reward and typically a team reward is considered with the form of $r(\bm{s}, \bm{a}): \mathcal{S} \times \mathcal{A}^N \rightarrow \mathbb{R}$. Partial observability is assumed in this scenario, as each agent has its own individual observations $\bm{z} \in \mathcal{Z}$ according to the observation function $O(\bm{s}, n): \mathcal{S} \times \mathcal{A} \rightarrow \mathcal{Z}$. Each agent stores information about past actions and observations using $\tau_n \in \mathcal{T}:= (\mathcal{Z} \times \mathcal{A})^*$, on which it conditions a stochastic policy $\pi_n(a_n|\tau_n): \mathcal{T} \times \mathcal{A} \rightarrow [0,1]$. Finally, a joint policy $\pi$ of the system presents a joint action-value function $Q^{\pi}_{jt}(s_t, \bm{a}_t) = \mathbb{E}_{s_{t+1}:\infty,\bm{a}_{t+1:\infty}}[R_t|s_t,\bm{a}_t]$, where $R_t = \sum_{i=0}^\infty \gamma^i r_{t+1}$ corresponds to the discounted return. All the multiagent RL algorithms utilized in this work, use the centralized training and decentralized execution (CTDE) paradigm that has shown more stability than independent learning for many tasks \cite{lowe2017multi}.

\subsection{Value Function Factorization}
Value Function Factorization (VFF) in Cooperative MARL refers to the decomposition of a joint action-value function into individual value functions or components that can be more easily managed and computed. This is particularly useful since the overall team complexity can be approximated with the contributions of each agent. In the introductory paper\cite{sunehag2017value}, the authors introduce the assumption that the joint action-value function could be approximated or factorized in less complex functions as, 
\begin{equation}
\label{vdn_fact}
    Q^{\pi}_{jt}(s_t, \bm{\tau}_t, \bm{a}_t) = \sum_{n=0}^N Q_n(s_t, \tau_{t,n}, a_{t,n}),
\end{equation}
where $Q_n(s_t, \tau_{t,i}, a_{t,i})$ corresponds to the action-value function of the $n^{th}$ agent. This assumption is known as \textit{Value Function Factorization}. A wider definition is introduced in \cite{son2019qtran} named the Individual-Global-Max (IGM). IGM defines that any MARL algorithm that can factorize agents' sequential decisions in the centralized training state holds the following, 
\begin{equation}
    \label{igm}
    \text{argmax}_{\bm{a}} Q^{\pi}_{jt}(s_t, \bm{\tau}_t, \bm{a}_t) = \begin{pmatrix}
     \text{argmax}_{a_1} Q_1(s_t, \tau_{t,1}, a_{t,1}) \\
       \vdots \\
     \text{argmax}_{a_N} Q_N(s_{t,N}, \tau_{t,N}, a_{t,N})
\end{pmatrix}    
\end{equation}
Equation \eqref{igm} is considered as a general definition of \eqref{vdn_fact}, as \eqref{vdn_fact} is only valid under the assumption of full factorization, while \eqref{igm} encompasses a larger set of joint Q-function families. Equation \eqref{vdn_fact} assumes that the joint Q-function is the addition of all individual agents' Q-function which leads to joint action selection while \eqref{igm} assumes that the joint action selection corresponds to the collection of all individual agents' actions. 

Numerous algorithms have been proposed based on the aforementioned assumption. However, due to scope limitations, we focus solely on those we modify by incorporating explainability through reward decomposition. Apart from Value Decomposition Networks (VDN) \cite{sunehag2017value}, introduced in \eqref{vdn_fact}, two algorithms have demonstrated state of the art performance: QMIX (Mixture of Q-values) \cite{rashid2020monotonic} and QTRAN (Transformation of Q-values) \cite{son2019qtran}. Differently from VDN, the authors of QMIX assume monotonicity via $\frac{\partial Q_n(s_t, \bm{\tau}_t, \bm{a}_t)}{\partial Q^{\pi}_{jt}(s_{t,n}, \tau_{t,n}, a_{t,n})} \geq 0, \forall n \in \mathcal{N}$. In the case of QTRAN, the authors identify some limitations in VDN and QMIX and propose a factorizable joint Q-function. This joint Q-function corrects inherent discrepancies due to partial observability by adding a compensation term to the summation of individual Q-functions. The additional term corresponds to a state-value function $V(s_t, \bm{\tau}_t, \bm{a}_t)$ described in \cite[Theorem~1]{son2019qtran} as,
\begin{equation}
    \sum_{n=0}^N Q_i(s_t, \tau_{t,i}, a_{t,i}) - Q^{\pi}_{jt}(s_t, \bm{\tau}_{t}, \bm{a}_{t} ) + V(s_t, \bm{\tau}_t, \bm{a}_t) = \begin{cases}
        0 & \bm{a} = \Bar{\bm{a}} \\
        \geq 0 & \bm{a} \neq \Bar{\bm{a}},
    \end{cases}
\end{equation}
where, 
\begin{equation}
    V(s_t, \bm{\tau}_t, \bm{a}_t) = \text{max}_{\bm{a}}Q^{\pi}_{jt}(s_t, \bm{\tau}_{t}, \bm{a}_{t} ) - Q_i(s_t, \tau_{t,i}, \Bar{a}_{t,i}).
\end{equation}
In the next subsection, we introduce Explainable Reinforcement Learning concepts and reward decomposition. 

\subsection{Explainable Reinforcement Learning (XRL) }
Explainable Reinforcement Learning is a recent subfield of Explainable ML that aims to shed light on the interpretation and explanations of RL algorithms. Unlike Supervised Learning (SL) and Unsupervised Learning (USL), RL is a sequential decision-making approach where an agent takes actions and seeks to maximize an objective function in the long term using a predefined exploration strategy \cite{sutton2018reinforcement}. Yet, the rationale behind the agent's actions is not always understood \cite{qing2022survey}. In \cite{milani2024explainable} a comprehensive taxonomy is described for XRL. The authors categorize XRL techniques into three groups based on the aspect of the RL agent they aim to explain: feature importance, which seeks explanations for individual actions; learning process and MDP, which finds explanations for influential experiences; and policy-level, which provides insights into long-term behavior. \footnote{Due to out of scope reasons, we do not describe the defined techniques in the first and third group of \cite{milani2024explainable}.}. In this work, we explore a technique found in the second group of the previous taxonomy, Decomposed Reward Functions (DRF). The main concept of this technique involves deriving insights from the agent's reward function by decomposing the composite objective function into separate individual objectives \cite{juozapaitis2019explainable}. Formally, a single agent composite reward function can be decomposed in $C$ components as,
\begin{equation}
    R(s_t, a_t) = \sum_{c\in C}R_c(s_t,a_t),
\end{equation}
where $R_c(s_t, a_t)$ corresponds to an individual objective for state $s_t$ and action $a_t$ in time $t$. Similarly, the authors in \cite{juozapaitis2019explainable} claim that $Q^{\pi}$ can be also decomposed as follows,  
\begin{equation}
    Q^{\pi}(s_t, a_t) = \sum_{c\in C}Q^{\pi}_c(s_t,a_t),
\end{equation}
where $Q^{\pi}_c(s_t,a_t)$ is the decomposed Q-value function corresponding $R_c$. In the next subsection, we introduce reward decomposition for Cooperative VFF-based MARL.

\subsection{Reward Decomposition for Cooperative Multi-Agent Reinforcement Learning}

Value Function Factorization algorithms such as VDN, QMIX, and QTRAN can also be described using the Multi-Agent Markov Games (MAMG)\cite{littman1994markov} framework as demonstrated in \cite{dou2022understanding}. This framework allows the analysis of reward decomposition on multi-agent valued decomposition algorithms concerning the Dec-POMDP framework since it can be considered a generalized case for Dec-POMDP. In addition, in \cite{macglashan2022value} some efforts have been made to study Q-function recovery in single agent domains. Inspired by the latter two works, we combine reward decomposition and multi-agent systems and propose Theorem \ref{theorem}.

\begin{theorem}\label{theorem}
\textit{Let a multi-agent Markov Game $\mathcal{M}\mathcal{G}(\mathcal{S}, \mathcal{A}, P, r, N, \gamma, d_0)$ with an individual agent decomposed objective function $r(s,a)$ be,
\begin{equation}
    r(s,a) \stackrel{\text{def}}{=} \sum_{c=1}^{C}\sigma_{c}r_{c}(s,a),
\end{equation}
where $\sigma_{c}\in [0,1], \sum_{c\in C}\sigma_c = 1$ corresponds to the weight for the $c^{th}$ decomposed reward and $r_{c}(s,a)\rightarrow \mathbb{R}$ the $c^{th}$ reward component for the state-action pair $s$ and $a$. Let the corresponding $i^{th}$ agent Q-value function $Q_i^{\pi}$ for the $c$ decomposed reward under policy $\pi$ be, 
\begin{equation}
    Q_i^{\pi} \stackrel{\text{def}}{=} \mathbb{E}_{\pi}\left[\sum_{t=0}^\infty \gamma^t r_i(s_t, a_t)| s_0 = s, a_0 = a \right].
\end{equation}
Consequently, a decomposed joint state-action Q-value function takes the form, 
\begin{equation}
    \label{theorem_eq}
   \bm{Q}^{\bm{\pi}}(s, \bm{a}) \stackrel{\text{def}}{=}   \sum_i^N  \sum_c^C  \sigma_{c,i}Q_{c,i}^{\pi}.
\end{equation}
}

\end{theorem}

A proof of Theorem \ref{theorem} is provided in Appendix \ref{Section8}. The theorem demonstrates that the joint action-value function can be decomposed in MARL. In addition, we empirically corroborate that performance is not affected in the next sections. Additionally, recent theoretical advances in the analysis of convergence in valued decomposition algorithms such as in \cite{wang2021towards} can guarantee the convergence of this family of algorithms as long as the decomposition from \eqref{theorem_eq} is valid.

\subsection{Explainable Reinforcement Learning Metrics} \label{explainable_metrics}
Capturing explanations or insights into the behavior of reinforcement learning (RL) algorithms has been a topic of research within the AI community. In the context of reward decomposition, various metrics have been proposed in the literature. These metrics aim to explain the behavior in terms of action preference for each decomposed reward. This is quite useful due to the direct implications for understanding why one algorithm outperforms another, why a particular action is preferred over others, and which reward is the most significant at the time of action decision. In \cite{juozapaitis2019explainable} three metrics are introduced, 

\begin{itemize}
    \item \textbf{Reward Difference Explanations (RDX)}: It is defined as $\Delta_c (s, a_x, a_y) = \bm{Q}_c(s, a_x) - \bm{Q}_c(s,a_y)$ and tries to provide insights about why an agent prefers action $a_x$ over $a_y$. The computed RDX can be plotted to show graphically the action preference per decomposed reward.   
    \item \textbf{Positive and Negative MSX (MSX$^+$ and MSX$^-$)}: This metric aims to provide concise graphical explanations by identifying the minimal number of rewards needed to select an action as RDX dimensions increase with $C$. It introduces the concept of the action disadvantage of $a_x$ over $a_y$ by defining $d:= \sum_{c\in C} I[\Delta_c (s, a_x, a_y) < 0] \cdot |\Delta_c (s, a_x, a_y)|$, where $I$ corresponds to the identity function. Then, two cases are possible: the first comprises the smallest set of positive values $X^+ \in \mathcal{X}$, and the second comprises the smallest set of negative values, $X^- \in \mathcal{X}$, calculated by $d$ for all $a_x$ to outweigh or to fall behind $a_y$, identified as MSX$^+$ and MSX$^-$, respectively.  MSX$^+$ can formally be written as, 
    \begin{equation}
        \text{MSX}^+ := \text{argmin}_{X^+ \in 2^C}|X^+| \text{s.t} \sum_{c\in C} \Delta_c (s, a_x, a_y) > d.
    \end{equation}
    
    \text{MSX}$^-$ is calculated differently since it represents the disadvantages of choosing $a_x$ over $a_y$. The ``just insufficient" advantage is calculated as: 
    \begin{equation}
        v := \sum_{c \in \text{MSX}^+} \Delta_c (s, a_x, a_y) - \text{min}_{c \in \text{MSX}^+ } \Delta_c (s, a_x, a_y), 
    \end{equation}
    then, $\text{MSX}^-$ can be defined as, 
    \begin{equation}
         \text{MSX}^-:= \text{argmin}_{X^- \in 2^C}|X^-| \text{s.t} \sum_{c\in C} -\Delta_c (s, a_x, a_y) > v
    \end{equation}
\end{itemize}

\subsection{Impact of delay, jitter and loss in QoE for XR services}
To the best of our knowledge, there are no recommendations or standards that include the impact of general Key Performance Indicators (KPIs) on XR services throughout the literature. However, some efforts have been made. In the recommendation ITU-T G.1035 \cite{itug1035}, three Virtual Reality (VR) QoE influence factors are depicted. These factors are categorized as follows:
\begin{enumerate}
\item \textbf{Human}: This category is related to human subjectivity, such as visual or hearing impairments, immersion capacity, demographics, or emotions-related factors.
\item \textbf{System}: This category is related to the content, video/audio codecs such as bitrate, frame rate, resolution, hardware, and network transmission.
\item \textbf{Context}: This category is related to those aspects that describe a situational influence on the user experience.
\end{enumerate}
Network transmission factors are related to the impact of networking indicators such as delay, jitter, or packet loss on VR. In the case of delay, including queuing, over-the-air, and buffering delays, it is one of the main causes of user discomfort due to simulator sickness \cite{9123114}. Packet loss corresponds to another important factor in the degradation of QoE \cite{8917613}. High packet losses in reliable transmission protocols increase the number of retransmissions and consequently the network delay. Jitter is described as the variation of the delay time of consecutive packets. The typical reason for large jitter corresponds to network congestion. A large amount of jitter can affect the user perception, especially in voice-related \cite{9642000}, XR \cite{akyildiz2022wireless} and haptic \cite{promwongsa2020comprehensive} applications.  The rest of the factors are not discussed in this work due to being out of scope. Despite 5G having increased considerably the reliability, bandwidth, and reduced latency of its plethora of services, the current XR codec adaptation state-of-the-art algorithms struggle in extreme situations where congestion or environmental losses exist\cite{iturria, Bojovic2023}. 

\section {System Model} \label{section4}

In this paper, we apply explainability to an application, specifically XR codec adaptation over 5G. To this end, we begin by providing the system model as follows. We utilize a 5G network accommodating a group of $N$ XR and CG users receiving downlink traffic from a Base Station (BS). The XR data flow originates from an application server, which periodically receives feedback from the BS at intervals of $t_w$. This feedback loop involves the aggregation of Key Performance Indicators (KPIs) from both the BS and the users, facilitating an exchange of information with the application server. With each feedback iteration, the application server dynamically adjusts XR codec parameters to uphold QoE standards. Our experimental setup features three distinct XR users connected to a single BS. To mimic real-world network conditions, we introduce adjustments to certain simulation parameters, such as the capacity of the transmission Radio Link Control buffer (Unacknowledged Mode) and the available bandwidth. 

Additionally, User Equipments (UEs) are spatially distributed across three concentric rings. Notably, the generated XR and CG traffic adheres to the specifications outlined in the 3GPP release-17 study \cite{3gpp3}.

\section{XR Codec Adaptation-based Explainable Multi-Agent Reinforcement Learning } 
In this section, we introduce the details and considerations in the design of the XR Codec Adaptation MARL algorithms, including the Markov Decision Process (MDP) components and architectural changes to VFF-based algorithms.

\begin{figure*}[t]
\center
  \includegraphics[scale=0.73]{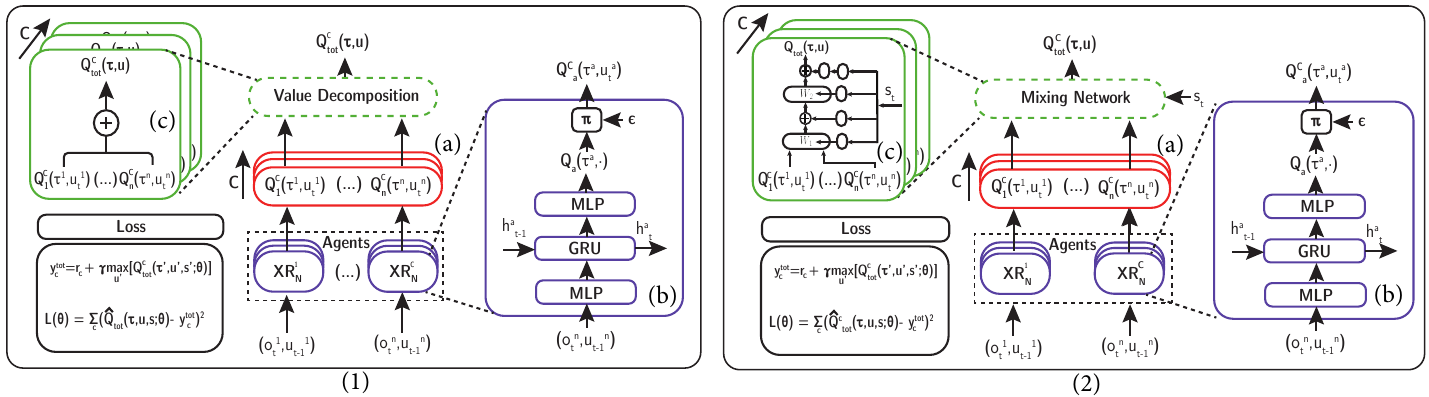}
  \setlength{\belowcaptionskip}{-5pt}
  \caption{Internal structure of the adapted explainable MARL algorithms: $\bm{(1)}$  Overview of the Decomposed Value Decomposition Networks (DVDN):  $\bm{(a)}$ $C$ Individual action-value networks $\bm{(b)}$ Details of the individual action-value network $\bm{(c)}$ $C$ additive summation of Q-values. $\bm{(2)}$ Overview of the Decomposed QMIX (DQMIX, Mixture of Q-Values): $\bm{(a)}$ $C$ Individual action-value networks $\bm{(b)}$ Details of the individual action-value network $\bm{(c)}$ $C$ mixing networks. } 
  \label{DVDN_DQMIX}

\end{figure*} 

\begin{algorithm}[h!]

\scriptsize 
	\caption{Decomposed VDN }
	\label{dvdn_algo}
	\begin{algorithmic}[1]
		\STATE Let $\mathbf{Q}_C = \{Q_1, ... , Q_c\}$ and $\mathbf{Q}_{tot} = \{Q_{tot,1}, ... , Q_{tot,c}\}$ be a vector comprised by $Q_c$ and $Q_{tot, c}$ , representing the Q-function and the mixing networks for each $c$ decomposed reward, respectively.  Initialize $\theta_c$ and $\theta'_c$ indicating the target and evaluation recurrent policies for the $c$ reward. Set the learning rate $\alpha$ and replay buffer $\mathcal{D} = \left\{ \right\}$.
		\STATE $\text{episode} = 0, \theta^{'}_c = \theta_c$  
       
		\FOR{environment episode $e\gets1$ \textbf{to} $E$}
		\STATE{$t = 0, o_0=\text{initial state}$}
		\WHILE{$done \neq True $ and $t < \text{steps/episode limit}$}
		
        \FOR{each $Q_c$ in $\mathbf{Q_C}$}
            \FOR{each agent $m$}
        		\STATE $\tau^m_t = \tau^m_{t-1} \cup \{(o_t, u_{t-1})\}$
        		\STATE $\epsilon = \text{epsilon-schedule}(\text{step})$
        
                \IF {$\epsilon \geq \mathcal{N}(0,1)$}
                \STATE{$randint(1, |U|^m); |U|^m \subset \mathbb{Z}^+$}
                \ENDIF                        
                \STATE {$\hat{\mathbf{Q}}^m_{c,t} \cup Q_c(\tau^m_t, u^m_t)$}  
    		\ENDFOR

        \ENDFOR
        \STATE {$u^m_t=             
            \begin{cases}
    			\argmax_{u^m_{c,t}}\sum_c \hat{\mathbf{Q}}^m_{c,t}~~~~\text{with probability }1 - \epsilon\\
    			randint(1, |U|^m); |U|^m \subset \mathbb{Z}^+ ~~ \text{with probability }\epsilon\\         
    	\end{cases}$}  
        \STATE {$\mathbf{u}_t = \{u^1_t, ..., u^m_t\}$}
  
		\STATE Get reward $r_t$ and next state $s_{t+1}$
		\STATE $\mathcal{D} = \mathcal{D} \cup \left\{(s_t, \mathbf{u}_t, r_1, ..., r_c, s_{t+1} )\right\} $
		\STATE{$\text{steps}=\text{steps}+1$}

		\ENDWHILE
  
		\IF {$|\mathcal{D}| > \text{batch-size}$}
		\STATE {$b \leftarrow$ random batch of episodes from $\mathcal{{D}}$}
		\FOR {each timestep $t$ in each episode in batch $b$}
		\STATE {$Q_{tot,c} = \sum_n Q_{n,c}(\tau^n_t,u_t^n)$}

		\ENDFOR
		
  \STATE {$\mathcal{L}_c(\theta_c)=\sum\limits_{i=1}^b\left[\left(y_i^{tot,c} - Q_{tot,c}(\boldsymbol{\tau}, \mathbf{u}, s; \theta_c) \right)^2\right]$}
		\STATE {$\Delta \theta_c = \nabla\theta_c(\mathcal{L}_c(\theta_c))^2$}
		\STATE {$\theta_c = \theta_c - \alpha \Delta \theta_c$}
		\ENDIF
		\IF {hard update steps have passed}
		\STATE {$\theta^{'}_c = \theta_c$} 
		\ENDIF
		\ENDFOR
  
	\end{algorithmic}
\end{algorithm}

\begin{algorithm}[h!]

\scriptsize 
	\caption{Decomposed QMIX }
	\label{dqmix_algo}
	\begin{algorithmic}[1]
		\STATE Let $\mathbf{Q}_C = \{Q_1, ... , Q_c\}$ and $\mathbf{Q}_{tot,C} = \{Q_{tot,1}, ... , Q_{tot,c}\}$ be a vector comprised by $Q_c$ and $Q_{tot, c}$ , representing the Q-function and the mixing networks for each $c$ decomposed reward, respectively. Initialize $\theta_c$ and $\theta'_c$ indicating the target and evaluation recurrent policies for the $c$ reward, the parameters of mixing network, agent networks, and hypernetwork. Set the learning rate $\alpha$ and replay buffer $\mathcal{D} = \left\{ \right\}$.
		\STATE $\text{episode} = 0, \theta^{'}_c = \theta_c$  
       
		\FOR{environment episode $e\gets1$ \textbf{to} $E$}
		\STATE{$t = 0, o_0=\text{initial state}$}
		\WHILE{$done \neq True $ and $t < \text{steps/episode limit}$}
		
        \FOR{each $Q_c$ in $\mathbf{Q_C}$}
            \FOR{each agent $m$}
        		\STATE $\tau^m_t = \tau^m_{t-1} \cup \{(o_t, u_{t-1})\}$
        		\STATE $\epsilon = \text{epsilon-schedule}(\text{step})$
        
                \IF {$\epsilon \geq \mathcal{N}(0,1)$}
                \STATE{$randint(1, |U|^m); |U|^m \subset \mathbb{Z}^+$}
                \ENDIF                        
                \STATE {$\hat{\mathbf{Q}}^m_{c,t} \cup Q_c(\tau^m_t, u^m_t)$}  
    		\ENDFOR

        \ENDFOR
        \STATE {$u^m_t=             
            \begin{cases}
    			\argmax_{u^m_{c,t}}\sum_c \hat{\mathbf{Q}}^m_{c,t}~~~~\text{with probability }1 - \epsilon\\
    			randint(1, |U|^m); |U|^m \subset \mathbb{Z}^+ ~~ \text{with probability }\epsilon\\         
    	\end{cases}$}  
        \STATE {$\mathbf{u}_t = \{u^1_t, ..., u^m_t\}$}
  
		\STATE Get reward $r_t$ and next state $s_{t+1}$
		\STATE $\mathcal{D} = \mathcal{D} \cup \left\{(s_t, \mathbf{u}_t, r_1, ..., r_c, s_{t+1} )\right\} $
		\STATE{$\text{steps}=\text{steps}+1$}

		\ENDWHILE
  
		\IF {$|\mathcal{D}| > \text{batch-size}$}
		\STATE {$b \leftarrow$ random batch of episodes from $\mathcal{{D}}$}
		\FOR {each timestep $t$ in each episode in batch $b$}
		\STATE {$Q_{tot,c} = \textit{Mixing-network~}(Q_{1,c}(\tau^1_t,u_t^1),\dots,Q_{n,c}(\tau^n_t,u_t^n)$;}  
           \STATE {\textit{Hypernetwork}$(s_t; \theta_c))$}
		\STATE {Calculate target $Q_{tot,c}$ using $\textit{Mixing-network}$ with $\textit{Hypernetwork}(s_t; \theta^{'}_c))$}
		\ENDFOR
		
  \STATE {$\mathcal{L}_c(\theta_c)=\sum\limits_{i=1}^b\left[\left(y_i^{tot,c} - Q_{tot,c}(\boldsymbol{\tau}, \mathbf{u}, s; \theta_c) \right)^2\right]$}
		\STATE {$\Delta \theta_c = \nabla\theta_c(\mathcal{L}_c(\theta_c))^2$}
		\STATE {$\theta_c = \theta_c - \alpha \Delta \theta_c$}
		\ENDIF
		\IF {hard update steps have passed}
		\STATE {$\theta^{'}_c = \theta_c$} 
		\ENDIF
		\ENDFOR
  
	\end{algorithmic}
\end{algorithm}

\subsection{State space selection}
The state space of the $n^{th}$ agent within the MARL framework is presented as,
\begin{equation}
    s_{t}^n = \{\bm{a}_{(t-1)}, b_{t}^{RLC}, p_{t}^{XR}\},
\end{equation}
where $\bm{a}_{(t-1)}$ signifies a vector encompassing the team's preceding codec parameter selection, $b_{t}^{RLC}$ denotes the RLC buffer occupancy ratio at the base station and $ p_{t}^{XR}$ corresponds to the Packet Delivery Ratio (PDR)\footnote{PDR = 1 - PLR, where PLR corresponds to the Packet Loss Ratio.}. In each proposed algorithm, the first state component is used in mitigating partial observability through the application of a Gated Recurrent Unit (GRU) layer within the individual action-value function.

\subsection{Action space selection}\label{action_space}
The action space for the $n^{th}$ agent is defined as, 
\begin{equation}
    a_{t}^{XR} = \{d_{min}^{XR}, d_{min}^{XR} + \frac{d_{max}^{XR} - d_{min}^{XR}}{K_{o}-1},..., d_{max}^{XR}\},
\end{equation}

where $d_{min}^{XR}$ and $d_{max}^{XR}$ are the maximum and minimum values of the codec data rate values per XR traffic, respectively. The maximum and minimum values can be found in Table \ref{net_settings}.

\subsection{Reward function }
Previously in \cite{iturria}, the MDP formulation only considered, as an objective function, a metric called XR index introduced in \cite{Dou2021}. 
This metric evaluates the Packet Delivery Budget (PDB) and PDR, assigning them a scale from 1 to 5. Despite the successful results obtained in this work, this approach considered a composite reward, which did not offer a clear understanding of the individual contributions of each term. Following reward decomposition, we present three objective functions that measure performance during a predefined interval. The components considered in this work correspond to delay, PDR, and Data Rate Ratio (DRR). For each UE, the delay component can be defined as follows, 

\begin{equation}
    r_{d,t}^{XR} = 
    \begin{cases}
        1   & \text{if $d_t^{XR} \leq 7$ s,}\\
        0.75   & \text{if $7$ s $\leq  d_t^{XR}  \leq 10 $ s,}\\
        0.5   & \text{if $10$ s $\leq  d_t^{XR} \leq 13$ s,}\\
        0.25   & \text{if $13$ s $\leq  d_t^{XR}  \leq 20$ s,}\\
        0   & \text{otherwise,}\\
    \end{cases}  \\
\end{equation}
where $d_t^{XR}$ corresponds to the average end-to-end delay measured at time $t$ in any $XR$ or $CG$ agent. From this point, we will refer to both types of agents as $XR$. The values chosen to define the different intervals (in seconds) follow Table \ref{Table1}. Following, we define the reward function that tracks PDR as, 
\begin{equation}
    r_{p,t}^{XR} = 
    \begin{cases}
        1   & \text{if $p_t^{XR} \geq 99\%$ ,}\\
        0.5   & \text{if $95\% \leq p_t^{XR} \leq 99\%$, }\\
        0   & \text{otherwise,}\\
    \end{cases}  \\
\end{equation}
where $p_t^{XR}$ corresponds to the average PDR measured at time $t$ in any $XR$ agent. Analogously as the delay-based component, $p_t^{XR}$ intervals use Table \ref{Table1} as reference. The average values of all metrics are calculated in intervals of $t_w$. Finally, we define a third component, that suggests that all agents increase their data rate. Such reward is defined as, 
\begin{equation}
    r_{th, t}^{XR} =  {th}_t^{XR}/{th}_{max}^{XR},  
\end{equation}
where ${th}_t^{XR}$ corresponds to the average throughput at time $t$ and ${th}_{max}^{XR}$, the maximum configuration value for the data rate in any $XR$ agent. Finally, a team reward is defined by component as,
\begin{align}
    \hat{r}_{d, t}^{XR} = \text{\textbf{min}}(\bm{r}_{d, t}^{XR})_{\{1,...,N\}},  \\
    \hat{r}_{p, t}^{XR} = \text{\textbf{min}}(\bm{r}_{p, t}^{XR})_{\{1,...,N\}},  \\ 
    \hat{r}_{th, t}^{XR} = \text{\textbf{min}}(\bm{r}_{th, t}^{XR})_{\{1,...,N\}},
\end{align}
where $\bm{r}_{d, t}^{XR}, \bm{r}_{p, t}^{XR}$ and $\bm{r}_{th, t}^{XR}$ are vectors comprised by $N$ UEs measured rewards of delay, PDR and DRR, respectively. 

\begin{table}[ht]
\scriptsize 
  \centering
  \caption{KPI Mapping for XR \cite{Dou2021}} 
  \label{Table1}
  \begin{threeparttable} 
  
  \begin{tabular}{c c }  
    \toprule
    XR Quality Index (XQI) & PDR $(\%)$, PDB (ms)\\ [1ex] 

\hline 
5&   $(99,7)$  \\ 
4 &  $(99,10)$ \\
3 &  $(95,13)$\\ 
2 &  $(95,20)$ \\
1 &  PDR $< 95$ or PDB $> 20$ \\ 
  \bottomrule
  \end{tabular}
   \end{threeparttable}
   \vspace{-2mm}
\end{table}
\subsubsection{Reward shaping in reward decomposition}
We observed that a reward definition based on intervals does not permit the provision of additional rewards when any of the measured KPIs improve within the same interval. Therefore, inspired by \cite{ng1999policy} we propose utilizing a modified potential reward-shaping strategy to harness any observed improvements. Each reward component $c$ is shaped as follows:
\begin{align}
    R_{c,t}^{XR} = \hat{r}_{c,t}^{XR} + F_c(x_{c,t}^{XR},\hat{x}_{c,t}^{XR}),
\end{align}
where $F_c$ is the potential function of the $c\in C$ reward component, $x \in \{d, p, th\}$ describing each measured KPI used in the calculation of the rewards (delay, PDR, DRR) and $x_{t}^{XR}$, $\hat{x}_{t}^{XR}$ correspond to the observed and the next observed KPI of $x$ type, respectively. Additionally,  $F_c$ can be described as, 
\begin{equation}
     F_c = \gamma_p \phi(\hat{x}_{c,t}^{XR},\mathbb{g}(\hat{x}_{c,t}^{XR})) - \phi(x_{c,t}^{XR}, \mathbb{g}(x_{c,t}^{XR})), 
\end{equation}
where $\mathbb{g}$ is a function that provides the goal given a measured KPI and  $\gamma_p$ a discount factor. Notice that the goal in this case corresponds to the upper bound of each reward-defined interval. Finally, the $\phi$ function is defined as, 

\begin{equation}  
\label{phi_def}
    \phi(x, y, z ) = 1 - \frac{|y - x|}{z_{(norm)}}   
\end{equation}

where $z$ is used as a normalizing factor. In this work, we assume $y = z_{norm}$, which makes \eqref{phi_def} equivalent to $\phi(x, y) = 1 - \frac{x}{y}$.  

\begin{figure*}[t]
\center
  \includegraphics[scale=0.74]{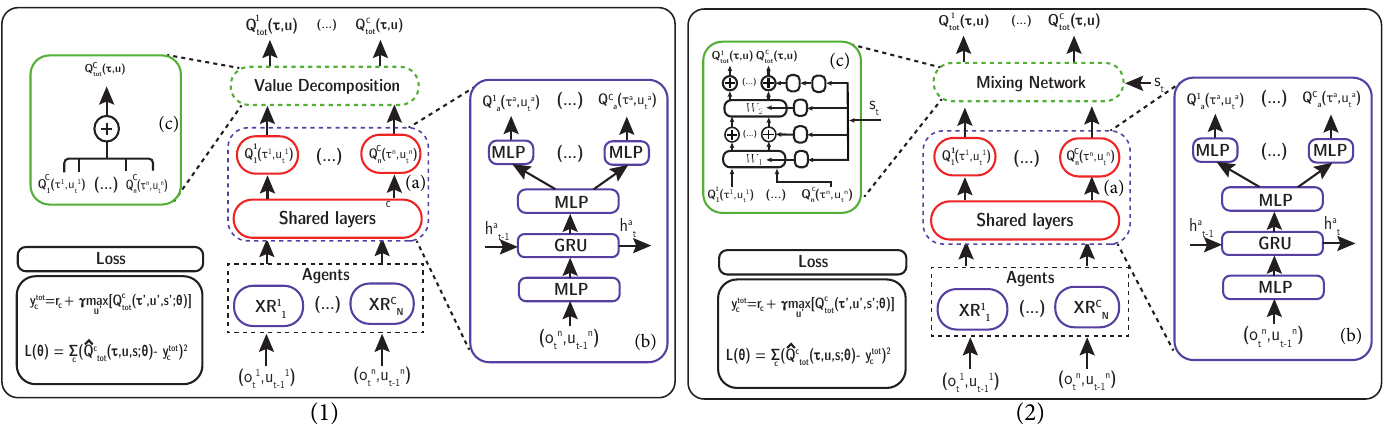}
  \setlength{\belowcaptionskip}{-5pt}
  \caption{Internal structure of the Multi-Headed Explainable MARL algorithms: $\bm{(1)}$  Overview of the Multi-Headed Decomposed Value Decomposition Networks (MH-DVDN): $\bm{(a)}$  Individual action-value network with shared layers $\bm{(b)}$ Details of the individual action-value network $\bm{(c)}$ Additive summation of Q-values. $\bm{(2)}$ Overview of the Multi-Headed Decomposed QMIX (MH-DQMIX, Mixture of Q-Values): $\bm{(a)}$  Individual action-value network with shared layers $\bm{(b)}$ Details of the individual action-value network $\bm{(c)}$ Mixing network.} 
  \label{dvdn_dqmix_mr}

\end{figure*} 

\subsection{Explainable Decomposed VDN Architecture}
VDN can be considered the first algorithm in the literature to introduce VFF in MARL. In Fig. \ref{DVDN_DQMIX} $\bm{(1)}$, the structure of the Decomposed Value Decomposition Networks (DVDN) is depicted. It comprises two main components: the mixing network and the agents' network. The mixing network involves an additive operation of all agents' Q-functions, as indicated by \eqref{vdn_fact}. In this scenario, no additional modifications are required for the previous operation, as it can be reused for each $c\in C$ reward component. Conversely, the network architecture for decentralized action execution changes with $C$ networks for each decomposed reward. Consequently, the loss function takes the following form:
\begin{equation}
\label{total_ltd_loss}
    \mathcal{L}(\bm{s}, \bm{u}, r, \bm{s}'; \bm{\theta} ) = \frac{1}{C}\sum_{c\in C} \mathcal{L}_{td}^c,
\end{equation}
where the temporal difference loss for the $c$ reward component is $\mathcal{L}_{td}^c = (Q_{tot}^c(\bm{s}, \bm{a}) - y_{tot}^c(r_c, \bm{s}; \bm{\theta}^-)^2$, $ y_{tot}^c(r_c, \bm{s}'; \bm{\theta}^-) = r_c + \gamma Q_{tot}^c(\bm{s}', \bm{\Bar{a}}';\bm{\theta}^-)$, $\bm{\Bar{a}}' = [\text{argmax}_{a_i}Q_i(s', a_i; \bm{\theta}^-)]_{i=1}^N$ and $\bm{\theta}'$ corresponds to the target network copied parameters from $\theta$. The full DVDN algorithm is presented in Algorithm \ref{dvdn_algo}.

\subsubsection{Reducing complexity on DVDN architecture}
As described previously, the proposed architecture complexity increases linearly as $C$ increases, which is undesirable due to the growing overhead in computational and time costs required for inference. To reduce the number of structures needed to accommodate the adaptation of reward decomposition into MARL, we propose modifying the structures of the individual value functions represented by a single agent network using inspiration from multi-task learning \cite{caruana1997multitask} as described in Fig. \ref{dvdn_dqmix_mr}$\bm{(1)}$. Each algorithm incorporating this improvement is prefixed with "Multi-Headed" (MH). For example, DVDN becomes MH-DVDN.

Comparatively to multi-task learning, each task becomes a Q-value function per $c\in C$ decomposed reward. From multiple structures, we reduce it to an individual neural network structure with shared layers and $C$ output heads, where each head predicts the Q-value for a single decomposed reward. During training, let's denote:

\begin{itemize}    
\item $\mathcal{D}$ as the replay buffer.
\item $\bm{b}_E$ as a random batch of episodes in tuple form $<s_t, \bm{a}_t, r_{1,t}, ..., r_{C,t}, s_{t+1}>_T$.
\item $r_{1,t}, ..., r_{C,t}$ correspond to the output data for the $C$ Q-value functions and are used to calculate the loss per decomposed reward.
\item $f$ as the shared neural network with parameters $\bm{\theta}$.
\item $g_1,..., g_C$ as the output heads with parameters $\bm{\kappa}_1, ..., \bm{\kappa}_C$, respectively.
\end{itemize}
The final structure of the network for the $i^{th}$ agent and $c^{th}$ decomposed reward  can be expressed as follows:

\begin{equation}
\label{final_structure_mr}
q_{i}^c = h(\bm{b}_E) = g_c(f(\bm{b}_E;\bm{\theta});\bm{\kappa}_c)
\end{equation}

Consequently, a loss function dependency changes to $\mathcal{L}_{td}^c(\bm{s}, \bm{u}, r_c, \bm{s}'; \bm{\theta}, \bm{\kappa}_c  )$. In the same fashion, $\mathcal{L}_{td}^c$ is used in \eqref{total_ltd_loss} to calculate the total loss. Note that in practice, the loss calculation in \eqref{total_ltd_loss} is similar to that used in the multi-headed modification; however, these represent different adaptations. The first approach uses multiple structures and a linear addition of all reward components, while the second approach leverages a single structure inspired by multitask learning, with linear addition applied to the reward components represented by each output head.

\subsection{Explainable Decomposed QMIX Architecture}
As introduced in previous sections, QMIX enforces a non-negative monotonic behavior of the joint Q-value function. This is achieved by utilizing a mixing network that establishes a direct relationship between individual Q-value functions and the joint one. Consequently, if any agent's Q-value increases, the joint Q-value function also increases; if it remains constant, the joint Q-value function remains constant; and so forth. As depicted in Fig. \ref{DVDN_DQMIX} $\bm{(2)}$, in the Decomposed QMIX (Mixture of Q-Values) or DQMIX, we utilize $C$ mixing networks to enforce monotonicity for the $C$ reward decomposition. The full DQMIX algorithm is presented in Algorithm \ref{dqmix_algo}.
\subsubsection{Reducing complexity on DQMIX architecture}

In the case of DQMIX, we proceed similarly to DVDN. The individual value function networks used by each agent are replaced by a multi-headed shared network architecture, as shown in \eqref{final_structure_mr}. However, some changes need to be made to the mixing network, as depicted in Fig. \ref{dvdn_dqmix_mr}$\bm{(2)}$. To describe the mentioned modifications that resulted in the Multi-Headed QMIX (MH-QMIX), let's denote:

\begin{itemize}
\item $\bm{q}_i = <q_i^1,...,q_i^C>$ as the Q-values for the $i^{th}$ agent and its $C$ decomposed Q-values.
\item The new hypernetwork architecture is comprised of a shared structure connected to $C$ individual heads. The shared network $\phi_{sh} = h_{sh}(s;\psi_{sh})$ is used to provide the hidden weights $w_{h}^c$ to each individual head as in $\phi_{c} = h_{c}(s, w_{h}^c; \psi_{c})$.
\end{itemize}

The output of the mixing network provides the joint value function $\bm{Q}_{tot}^{i} = \{Q_{tot}^{i,1},..., Q_{tot}^{i,C}\}$ for agent $i$,

\begin{equation}
    \bm{Q}_{tot}^{i}(s, \bm{q}_i; \bm{\theta}, \bm{\psi}_{sh}, \bm{\psi}_{h}) = g_i(|\phi_{c}(s, \bm{w}_{h},|\bm{\phi}_{sh}(s; \bm{\psi}_{sh})| \cdot  \bm{q}_i; \bm{\psi}_{h}|),
\end{equation}
where the $\cdot$ operator corresponds to the matrix-matrix product, $g_i$ is the mixing network for agent $i$, $\bm{w}_{h}= \{\bm{w}_{h}^1,...,\bm{w}_{h}^C\}$ is a vector with the hidden weights per head provided by the shared network, $\bm{\psi}_{h} = \{\bm{\psi}_{h}^1,...,\bm{\psi}_{h}^C\}$ is a vector with the parameters of each head of the architecture. It is important to note that only weights are enforced to be non-negative leaving biases to acquire negative values.

\subsection{Explainable Decomposed QTRAN Architecture}
In this subsection, we delve into the details of the Decomposed Transformation of Q-Values or DQTRAN framework. Vanilla QTRAN presents three main blocks comprised of neural networks: (i) an individual action-value network $Q_n$, (ii) a joint action-value network $Q_{tot}$, and (iii) a state-value network $V_{tot}$. We proceed similarly as in DVDN and DQMIX and modify the structure to allow reward decomposition. DQTRAN requires $C$ number of each block defined in vanilla QTRAN. 
\begin{figure}[t]
\center
  \includegraphics[scale=0.70]{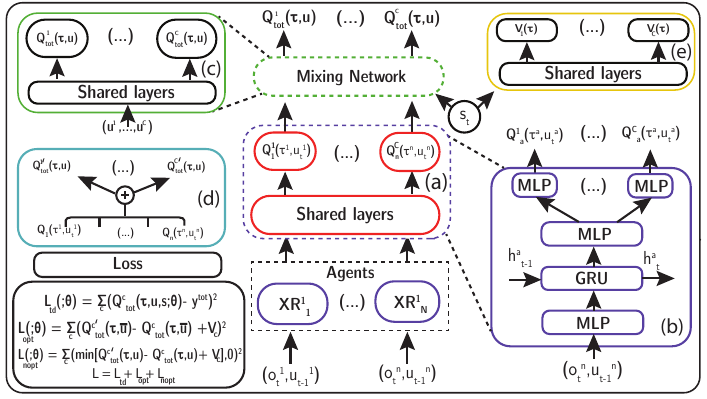}
  \setlength{\belowcaptionskip}{-5pt}
  \caption{Internal structure and loss calculation of the Multi-Headed Decomposed QTRAN (MH-DQTRAN): $\bm{(a)}$  Individual action-value network with shared layers $\bm{(b)}$ Details of the individual action-value network $\bm{(c)}$ Joint action-value network $\bm{(d)}$ Additive summation of Q-values $\bm{(e)}$ State-value network.} 
  \label{dqtran_architecture_mr}
\end{figure} 
Thus, in reward decomposition, the loss function of DQTRAN takes the form of, 
\begin{align}
    \mathcal{L}(\bm{s}, \bm{u}, r, \bm{s}'; \bm{\theta} ) = \frac{1}{C}\left(\sum_{c \in C}\mathcal{L}_{td}^c + \lambda_{opt}\sum_{c \in C}\mathcal{L}_{opt}^c + \lambda_{nopt}\sum_{c \in C}\mathcal{L}_{nopt}^c\right)
\end{align}
where $\lambda_{opt}$ and $\lambda_{nopt}$ are the weighting factors for $\mathcal{L}_{opt}^c$ and $\mathcal{L}_{nopt}^c$\footnote{Additional details of $\mathcal{L}_{opt}$ and $\mathcal{L}_{nopt}$ can be found in \cite{son2019qtran}}, losses involved in the factorization of the joint policy, respectively. 

\subsubsection{Reducing complexity on DQTRAN architecture}
Fig. \ref{dqtran_architecture_mr} presents the changes introduced to reduce the overhead in vanilla DQTRAN. In the same fashion as DVDN and DQMIX, the changes correspond to multi-headed structures with shared layers. Multi-Headed DQTRAN (MH-DQTRAN) consists of three multi-head function approximators, where each estimator has $C$ heads: (i) each agent \textit{n}'s action-value network for $Q_{n}^c$, (ii) a joint action-value network for $Q_{n,tot}^c$, and (iii) a state-value network for $ V_n^c$, \textit{i.e}, 

\begin{itemize}
    \item (\textbf{Individual action-value network})   $f_q^c :  (o_t^n, u_{t}^n) \rightarrow Q_{n}^c$
    \item (\textbf{Joint action-value network})   $f_r^c : (\bm{s_t}, \bm{u_t^n}) \rightarrow Q_{n,tot}^c$
    \item (\textbf{State-value network})   $f_v^c :\bm{s_t} \rightarrow V_n^c$
\end{itemize}

\subsection{Reward Difference Fluctuation Explanations (RDFX)}
In \cite{juozapaitis2019explainable}, a metric called RDX is introduced as described in subsection \ref{explainable_metrics}. However, such a metric can only be applied when actions are well-known. Typically, in wireless networks, actions consist of adjusting a value that can be continuous or whose dimension can become large. To overcome this situation, we propose a metric called Reward Difference Fluctuation Explanations (RDFX). It is defined as the difference between the chosen action or the maximum value of the $\bm{Q}_c$ vector and the positive or negative fluctuation of the rest of the actions in a given state. Formally, we can define it as:

\begin{equation}
    \Delta_c(s, \bm{Q}_c, ``fluct") = max(\bm{Q}_c) - Q_{c, fluct},
\end{equation}
where $``fluct"$ can have two meanings, a positive fluctuation $``inc"$ or negative fluctuation $``dec"$ and $Q_{c, fluct}$ corresponds to the expected value due to increasing or decreasing value of the actions according to $``fluct"$.  

The positive fluctuation set can be obtained as follows. $\bm{Q}_c$ is a vector comprising all Q-values of the $c$ reward component, $\bm{Q}_{c,inc} \subseteq \bm{Q}_c$. The first element of $\bm{Q}_{c,inc}$ starts at the $i^{th} + 1$ element of $\bm{Q}_c$ where $q_i = \text{argmax}(\bm{Q}_c)$. Then, $\bm{Q}_{c,inc} = \{q_{i+1}, q_{i+2}, ..., q_N\}$, where $q_N$ is the final element of $\bm{Q}_c$. Similarly as $\bm{Q}_{c,inc}$, a negative fluctuation is defined as  $\bm{Q}_{c,dec} =  \bm{Q}_{c} \setminus \bm{Q}_{c,inc}$. Thus,  $\bm{Q}_{c,dec} = \{q_{0}, q_{1}, ..., q_{i-1}\}$. Finally,  $Q_{c, fluct}$ can be defined as,

\begin{equation} \label{Q_fluctuation}
    Q_{c, fluct} :=
    \begin{cases}
     \mathbb{E}[\bm{Q}_{c,inc}] \text{ as } ``inc"  & \text{if } \bm{Q}_{c,inc} \neq \varnothing \text{ else } \bm{Q}_c[N] \\    
     \mathbb{E}[\bm{Q}_{c,dec}] \text{ as } ``dec"  & \text{if } \bm{Q}_{c,dec} \neq \varnothing \text{ else } \bm{Q}_c[0] \\ 
    \end{cases}  \\
\end{equation}

Given the previous adjustment of the RDX, it is possible to utilize $\text{MSX}^+$ and $\text{MSX}^-$ utilizing RDFX instead.

\subsection{Adaptive Explainable MARL via episodic gradient importance} \label{adaptive_back}
\begin{figure*}[t]
\center
  \includegraphics[scale=0.68]{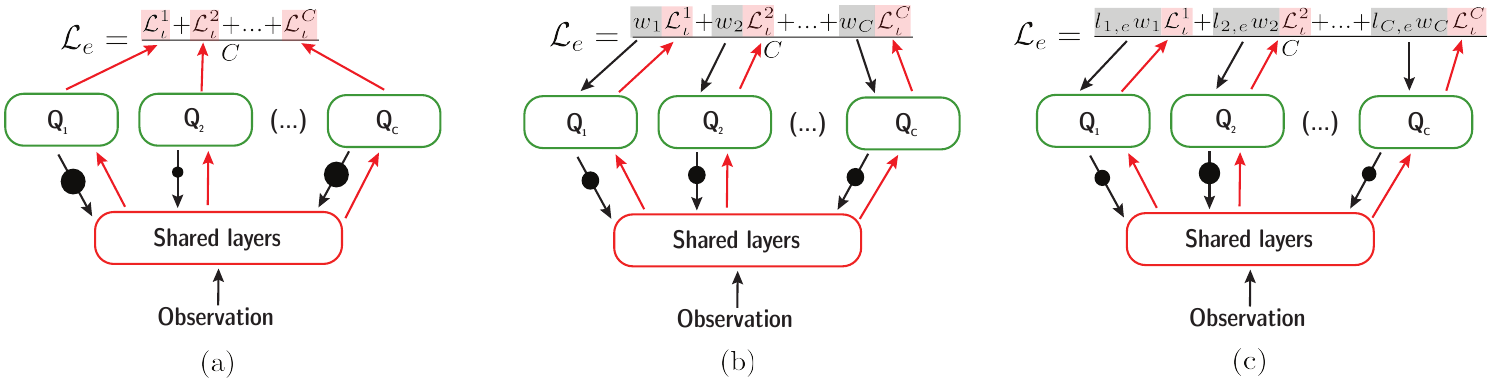}
  \caption{Summary of the importance gradient mechanism. $\bm{(a)}$ Multi-Headed loss function without gradient normalization $\bm{(b)}$ Multi-Headed loss function with gradient normalization $\bm{(c)}$ Multi-headed loss function with importance gradient.} 
  \label{importance_gradient}
\end{figure*}

In this work, we propose reducing the complexity of decomposed XMARL algorithms, drawing inspiration from multi-task learning. As mentioned previously, we modify the internal structures of each MARL algorithm to enable the outputting of independent Q-values per $c\in C$ reward. Reward decomposition is a technique that permits the analysis of reward components post-hoc, meaning that it's not possible to utilize observed reward behavior during training. In this subsection, we aim to propose a novel online algorithm that leverages reward decomposition to extract and utilize knowledge during training. The proposed algorithm draws inspiration from GradNorm \cite{chen2018gradnorm}, with a notable difference: while GradNorm normalizes training rates among tasks, our algorithm incentivizes differentiation among them. In our context, tasks are equivalent to decomposed Q-functions. Let's begin by describing some preliminaries, to propose our solution later. Fig. \ref{importance_gradient}$\bm{(a)}$ shows how gradients behave for each reward component Q-function output. The loss is calculated in the same fashion we introduced in \eqref{total_ltd_loss}. It can be seen that the gradient dimensions differ due to the data stored concerning each decomposed reward. To remove these inherent differences driven by the acquisition of observations, actions, and rewards into the experience replay, a new loss function is proposed to scale gradients to adjust the rate at which each decomposed Q-function must be trained, 
\begin{equation}
    \label{loss_gradnorm}
   \mathcal{L}_e=\frac{1}{C}\sum_{c \in C} w_{c} \mathcal{L}^{c}_{\iota}
\end{equation}
where $\mathcal{L}_e$ corresponds to the loss in episode $e$\footnote{Note that we considered the training to be performed in each episode.}, $w_{c}$ are the weights functions corresponding $c^{th}$ Q-function, $\mathcal{L}^{c}_{\iota}$ are the individual losses for each decomposed reward head and $\iota \in \{td, opt, nopt\}$. The previous weights adjust the training rate of each reward component during the training process. Now that the gradients are balanced, we utilize reward decomposition to enhance learning by assessing the significance of each reward component and incorporating it into the loss calculation. The new loss function is formulated as follows, 
\begin{equation}
    \label{loss_gradnorm1}
   \mathcal{L}_e=\frac{1}{C}\sum_{c \in C} w_{c} l_{c,e} \mathcal{L}^{c}_{\iota},
\end{equation}
where  $l_{c,e}$ are the reward importance weights for each decomposed reward head.
To obtain $l_{c,e}$, during each episode $e$, we collect a sequence of rewards per $c \in C$ reward type. To illustrate the statement, let's say we have three types of rewards, 
 \begin{equation}
 \begin{aligned} 
  \mathbf{r}_{1,e} &= \{r_{1,e}^1, ..., r_{1,e}^{ts}\}, \\ 
  \mathbf{r}_{2,e} &= \{ r_{2,e}^1, ..., r_{2,e}^{ts}\}, \\
   & \vdots\\
  \mathbf{r}_{c,e} &= \{ r_{c,e}^1, ..., r_{c,e}^{ts}\},
 \end{aligned}
 \end{equation}
 where $ts$ corresponds to the total of steps per episode. Next, we compute the means and variance of $r_1,...,r_c$ in episode $e$ using, 
\begin{equation}
  \bar{r}_{c,e} =  \frac{1}{|\mathbf{r}_{c,e}|}\sum_{t=1}^{ts} \mathbf{r}_{c,e}^t,
  \label{average_rew_episode}
\end{equation}
\begin{equation}
\sigma_{c,e} = \sqrt{\frac{1}{ts - 1} \sum_{t=1}^{ts} (\mathbf{r}_{c,e}^t - \bar{r}_{c,e} )^2}, 
\label{deviation_rew_episode}
\end{equation}
where $\bar{r}_{c,e}$ and $\sigma_{c,e}$ correspond to the mean and variance of the $c$ rewards, Finally, we define the reward weight $l_{c,e}$ by utilizing softmax such as $\sum_{c\in C} l_{c,e} = 1$. The proposed equation is, 
\begin{equation}
l_{c,e} = \frac{\text{exp}\left(\bar{r}_{c,e} + \sinh\left(\sigma_{c,e}\right) \right)}{\sum_{c\in C} \text{exp}\left(\bar{r}_{c,e} + \sinh\left(\sigma_{c,e}\right) \right).}
\label{weight_reward_episode}
\end{equation}

The selection of the previous equation solves an undesirable behavior observed in cases where the variance term is given no importance such as in, 
\begin{equation}
\label{no_variance_weights}
l_{c,e} = \frac{\text{exp}\left(\bar{r}_{c,e} \right)}{\sum_{c\in C} \text{exp}\left(\bar{r}_{c,e} \right)}.
\end{equation}
Each algorithm incorporating this improvement is prefixed with "Multi-Headed Adaptive" (MHA). For example, DVDN becomes MHA-DVDN.

\begin{algorithm}[h!]

\scriptsize 
	\caption{Adaptive gradient via decomposed reward importance}
	\label{adaptive_importance}
	\begin{algorithmic}[1]
		\STATE{Initialize $w_{c} = 1$  $\forall c$ and $l_{c,0} = 1$ $\forall c$. Select $\alpha_{\mathcal{W}} > 0$}
        \STATE{Initialize network parameters $\mathcal{W}$. According to the algorithm used this value will change, generally defined as the concatenation of all parameters of the involved neural networks.}
       
        \FOR{environment episode $e \leftarrow 1$ to $E$ do}
            
            \WHILE{$t < ts$}
                \IF {$|\mathcal{D}| > \text{batch-size}$}
        		  \STATE {b $\leftarrow$ random batch of episodes from $\mathcal{{D}}$}
                    \FOR {each timestep $t$ in each episode in batch $b$}
                		\STATE {$\mathcal{L}(t)=\frac{1}{C}\sum_{c \in C} w_{Q_c,e}(t) l_{c,e}(t) L_{c}(t)$}
                        \STATE {Compute $G^{\mathcal{W}}_{c}(t)$ and $t^r_{c}$ $\forall c$}
                        \STATE {Compute $\bar{G}^{\mathcal{W}}_{c}(t) = \frac{1}{C}\sum_{c \in C} G^{\mathcal{W}}_{c}(t)$}
                        \STATE {Compute $\mathcal{L}_{grad} = \sum_{c \in C} |G^{\mathcal{W}}_{c}(t) - \bar{G}^{\mathcal{W}}_{c}(t) \times [t^r_{c}]^{\alpha_{\mathcal{W}}}|_1$}
                        \STATE {Compute GradNorm gradients $\nabla_{w_{Q_c,e}}$}
                        \STATE{Compute gradients $\nabla_{\mathcal{W}}\mathcal{L}$}
                        \STATE{Update $w_{Q_C, e}(t) \rightarrow  w_{Q_C, e}(t+1)$} using $\nabla_{\mathcal{W}}\mathcal{L}$
                        \STATE{Normalize $w_{Q_C, e}(t+1)$}

                   \ENDFOR
                \ENDIF
            \ENDWHILE
            \STATE{Update $l_{c,e}(t) \rightarrow  l_{c,e}(t+1)$, using $\bar{r}_{c,e}$,  $\sigma_{c,e}$ and weights $l_{c,e}$ according to equations  (\ref{average_rew_episode}), (\ref{deviation_rew_episode}) and (\ref{weight_reward_episode}), respectively. }

        \ENDFOR

	\end{algorithmic}
\end{algorithm}
\subsection{Baselines: Adjust Packet Size Algorithm (APS) and proposed decomposed algorithms}
In this study, we compare our proposed schemes with an analog XR loopback threshold-based algorithm 
 named Adjust Packet Size (APS) delineated in \cite{Bojovic2023}. In the previous work, three algorithms are presented, making APS the best algorithm in terms of performance. 
Furthermore, we introduce variants of the algorithms VDN, QMIX, and QTRAN. Each of these variants is accompanied by a prefix indicating the type of decomposed modification it represents. For instance,
\begin{itemize}
    \item \textbf{Decomposed Vanilla}: Indicated by the letter \textbf{D} and described by figures \ref{DVDN_DQMIX}$\bm{(1)}$ and \ref{dvdn_dqmix_mr}$\bm{(1)}$.
    \item \textbf{Decomposed Multi-Headed}: Indicated by the letters \textbf{MH-D} and described by figures \ref{DVDN_DQMIX}$\bm{(2)}$, \ref{dvdn_dqmix_mr}$\bm{(2)}$, and \ref{dqtran_architecture_mr}.
    \item \textbf{Decomposed Multi-Headed Adaptive}: Indicated by the letters \textbf{MHA-D} and described by subsection \ref{adaptive_back}.
\end{itemize}
In the next section, we depict the results of the current work concerning explainability and network performance.  

\section{Performance evaluation}\label{section5}
In this section, we present the results of employing reward decomposition in MARL. We showcase the average Q-values, RDFX, and MSX metrics to facilitate a deeper understanding of the contribution of each reward component during action selection. These metrics enable enhancements for the proposed decomposed MARL algorithms. Additionally, we offer comprehensive comparative results among all the proposed algorithms in terms of network KPIs. We conduct simulations utilizing the ns-3 New Radio (NR) module, version 2.3 released in April 2023. Formerly known as NR-Lena, this module is integrated into the ns-3 simulator, offering simulations for 3GPP NR non-standalone cellular networks. Furthermore, we employ the ns-3 gym \cite{ns3gym} module as an interface between ns-3 and Python-based agents. 

\begin{figure}[t]
\center
  \includegraphics[scale=0.38]{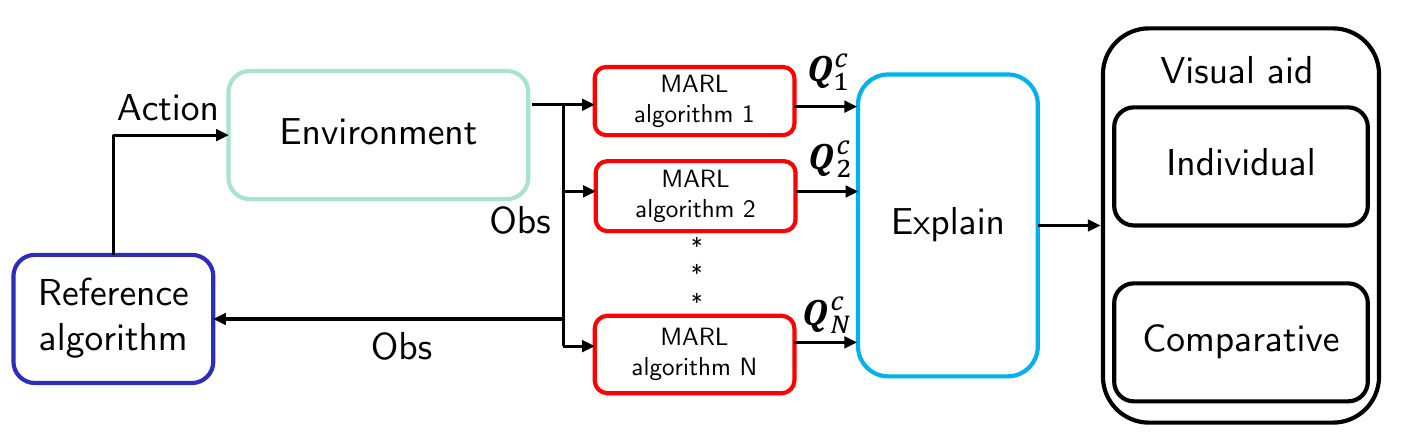}
 
  \caption{Explainability framework} 
  \label{explanaibility_framework}
\end{figure} 

\subsection{Simulation Settings}
RL parameters and simulation settings are delineated in Tables \ref{learning_settings} and \ref{net_settings}, respectively. To investigate the influence of distance on both the proposed algorithms and the existing baseline (APS), users are dispersed randomly across three distinct rings with radii ranging from 100 to 400 meters. Within each ring, users are randomly positioned within the inner radius and maintain random mobility throughout the simulation, moving at a speed of $3$ m/s. The capacity of the transmission RLC buffer is capped at 60,000 bytes. The simulation adopts an Urban Macro (UMa) channel model, with all UEs in non-line-of-sight conditions. It employs a single bandwidth part with a central frequency of 4 GHz and a bandwidth of 20 MHz. In the next subsection, we will discuss the details of the proposed evaluation framework.   

\begin{table}
\caption{Reinforcement Learning Settings.}

\begin{center}
\resizebox{\columnwidth}{!}{%
\begin{tabular}{c c} 
\hline
\textbf{Parameter}&\textbf{Value} \\
\hline

Maximum training steps & {$30\mathrm{e}{4}$} \\
Initial $\epsilon$ & {1} \\
$\epsilon$ decay steps & {$25e3$ steps} \\
$\epsilon$ minimum & {$5\mathrm{e}{-2}$} \\
Buffer $(\mathcal{D})$ size & {$5\mathrm{e}{3}$} \\
Batch size & {$32$} \\ 
Learning rate & {$8\mathrm{e}{-3}$} \\
Discount factor $(\gamma)$ & {$0.99$} \\
Recurrent Layer hidden dimension & {$64$} \\
MultiLayer Perceptron hidden dimension (QMIX) & {$32$} \\
MultiLayer Perceptron hidden dimension (QTRAN) & {$64$} \\
Reuse Network & {True} \\
Deep Q-Network structure & {Double Q-networks} \\
The number of layers of hyper-network  & {2} \\
The number of layers of head-network  & {2} \\
The dimension of the hidden layer of the hyper-network  & {64} \\
Reward shaping discount factor $(\gamma_p)$ & {1} \\
$\lambda_{opt}$ and $\lambda_{nopt}$ & {1, 1} \\
\hline
Number of agents in the MAS & {$3$}\\
Observation window ($t_w$) & {$0.5$ s}\\
\hline
\end{tabular}
}

\label{learning_settings}
\end{center}

\end{table}

\begin{table}
\caption{Network Settings}

\begin{center}
\resizebox{\columnwidth}{!}{%
\begin{tabular}{c c} 
\hline
\textbf{Parameter}&\textbf{Value} \\
\hline
Wireless network & { New Radio (NR) } \\
Channel Bandwidth & { $40$ MHz } \\
Central Frequency ($f_c$) & { $4$ GHz } \\
Number of UEs $(N)$ & { 3 } \\
Number of gNB & {$1$}\\
Propagation Loss Model & { UMa nLos }\\
Numerology & { 2 } \\
gNB Noise Figure & {$5$ dB}\\
gNB Transmission Power & {$43$ dBm}\\
gnB Antenna configuration & {4x8} \\
UE Noise Figure & {$7$ dB}\\
UE Transmission Power & {$26$ dBm}\\
UE Antenna configuration & {1x1} \\
Max Transmission Buffer Size & {60 KBytes} \\
XR and CG traffic characteristics & {AR (3 flows), VR (1 flow), CG (1 flow) } \\
Codec data rate [min,max] & {AR: [0.5,10], VR: [10,30], CG: [10,30] Mbps } \\
\hline
\end{tabular}
}
\label{net_settings}

\end{center}
\vspace{-3mm}
\end{table}

\subsection{Evaluation Framework}
In this subsection, we propose an evaluation framework (Fig. \ref{explanaibility_framework}) to facilitate the analysis of our proposed decomposed MARL algorithms. As illustrated in the figure, a reference algorithm (RA) - typically the baseline - and the decomposed MARL algorithm are initialized. This initialization involves loading the network parameters (weights and biases) of all algorithms to be evaluated. Subsequently, the RA interacts with the environment, and the observations generated based on the actions taken by the RA serve as input for the decomposed algorithms. At each time step, the Q-values from each algorithm are utilized to calculate the RDFX and MSX metrics. This process occurs in the Explain block (blue colored box in Fig. \ref{explanaibility_framework}), where the Q-values of each decomposed reward are analyzed using the previously proposed metrics. This process allows concurrent explanations of each algorithm's behavior and enables a fair comparison.

\subsection{Explanation insights for reward decomposition in XR codec adaptation}
In this subsection, we offer a quantitative evaluation of explainable metrics derived from employing the proposed explainable framework and the trained decomposed MARL algorithms. The primary objective is to illustrate how wireless-related metrics, frequently utilized in reward design, can influence the formulation of RL problems. Specifically, we utilize the average Q-value, RDFX, and MSX to draw conclusions regarding the components of a predefined composite function. Finally, we present the performance of the proposed decomposed algorithms and the baseline APS in terms of network KPIs.
\begin{figure*}[t]
\center
  \includegraphics[scale=0.61]{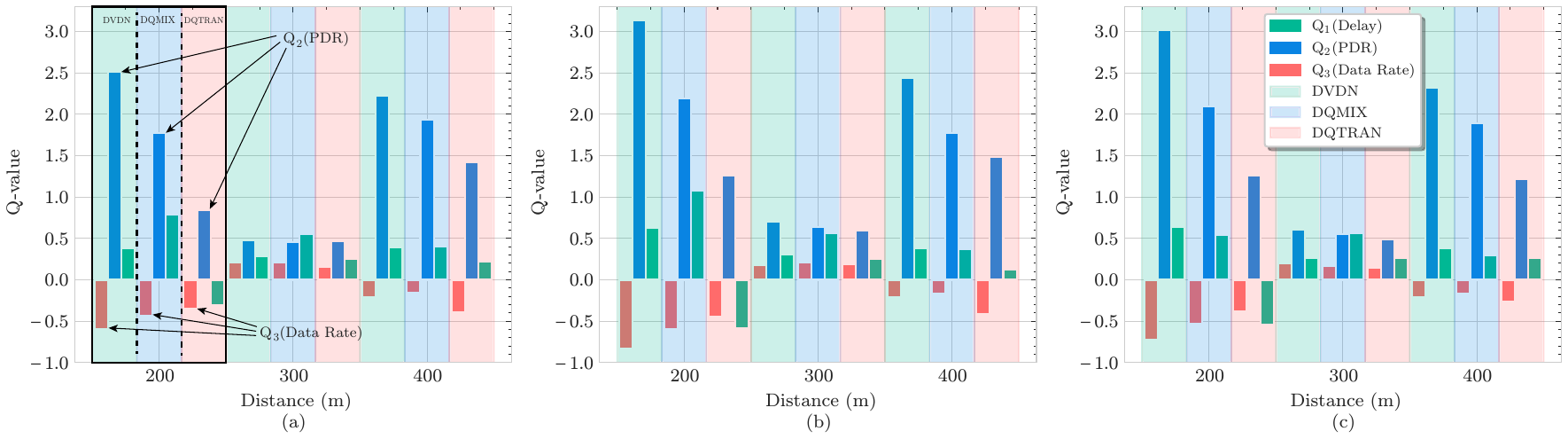} 
  \caption{Average Q-Value functions decompositions for DQTRAN, DQMIX, and DVDN for three distances $200, 300, 400$ m: $\bm{(a)}$ AR agent, $\bm{(b)}$ VR agent, and $\bm{(c)}$ CG agent. To aid readers, we have highlighted the Q-value information for $200$ m. Each colored column represents an algorithm type, and within each algorithm, the information for each reward is provided. Arrows indicate the highest Q-value, $Q_2$ (PDR), and the lowest, $Q_3$ (Data Rate). } 
  \label{q_value_graph}
\end{figure*}
\begin{figure*}[t]
\center
  \includegraphics[scale=0.61]{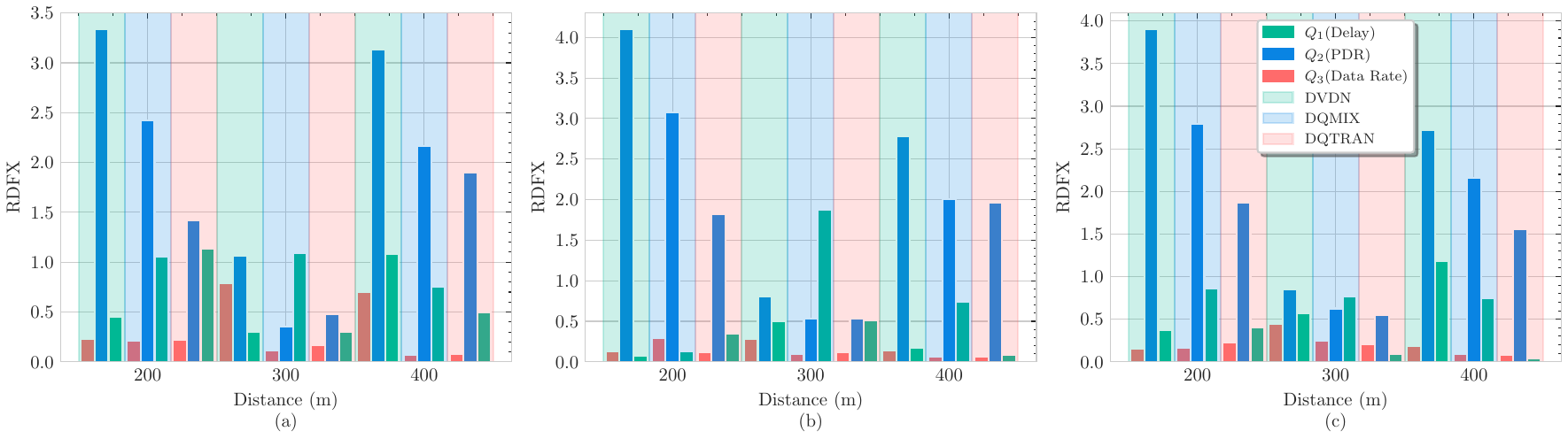} 
  \caption{Average RDFX $\Delta_c(\cdot, \bm{Q}_c, ``inc")$ for DQTRAN, DQMIX, and DVDN for three distances $200, 300, 400$ m: $\bm{(a)}$ AR agent, $\bm{(b)}$ VR agent, and $\bm{(c)}$ CG agent.} 
  \label{rdfx_inc}
\end{figure*}

\begin{figure*}[t]
\center
  \includegraphics[scale=0.61]{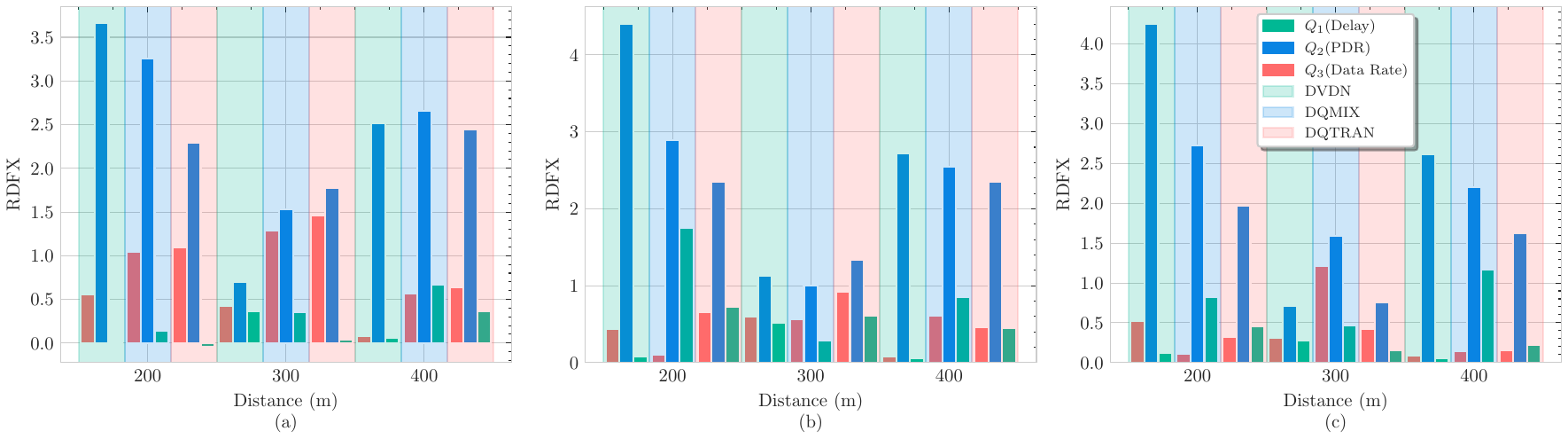} 
  \caption{Average RDFX $\Delta_c(\cdot, \bm{Q}_c, ``dec")$ for DQTRAN, DQMIX, and DVDN for three distances $200, 300, 400$ m: $\bm{(a)}$ AR agent, $\bm{(b)}$ VR agent, and $\bm{(c)}$ CG agent.} 
  \label{rdfx_dec}
\end{figure*}

In Fig. \ref{q_value_graph}, we display the decompositions of the Q-value functions for DQTRAN, DQMIX, and DVDN across distances of ${200, 300, 400}$ m and for the agents AR, VR, and CG, respectively. Each algorithm type is represented by a lighter color for each distance. Additionally, for each algorithm, three bar plots illustrate the average behavior of the Q-values for the selected action across reward components: $Q_1$, $Q_2$, and $Q_3$ for Delay, PDR, and DRR, respectively. To aid analysis, an example is given for 200 m in Fig. \ref{q_value_graph}$(\bm{a})$. Upon initial examination of Fig. \ref{q_value_graph}$(\bm{a})$, corresponding to the AR agent, it is evident that $Q_2$ (PDR) exhibits the highest Q-value, serving as the primary contributor in decision-making, followed by $Q_1$ (Delay). Similarly, we observe that $Q_3$ (Data Rate) has the lowest values (in some cases even negative, as in this example), indicating it as the least desirable reward. This pattern persists across all agents, distances, and algorithms, with one exception: the AR agent at 300 m with the QMIX algorithm, where the delay is slightly higher than the PDR. Despite this exception, we can conclude that, in general, PDR holds the highest importance.
Moreover, in Fig. \ref{rdfx_inc} and Fig. \ref{rdfx_dec}, we illustrate the average RDFX for two scenarios: selecting the current codec parameter versus selecting a higher codec parameter and selecting the current codec parameter versus selecting a lower codec parameter. Similar to Fig. \ref{q_value_graph}, we can observe that the RDFX is consistently positive for all actions chosen by each algorithm. According to the definition of RDFX, this signifies the superiority of the selected action compared to choosing a higher or lower value for the XR codec.

\begin{figure*}[t]
\center
  \includegraphics[scale=0.61]{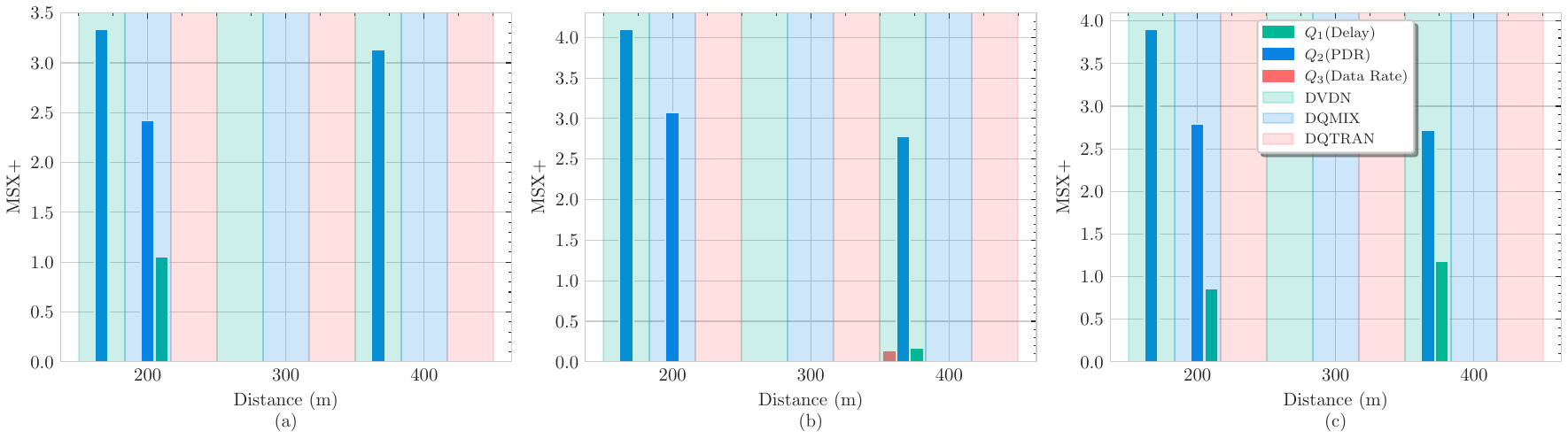} 
  \caption{Average MSX+$(``inc")$ for DQTRAN, DQMIX, and DVDN for three distances $200, 300, 400$ m: $\bm{(a)}$ AR agent, $\bm{(b)}$ VR agent, and $\bm{(c)}$ CG agent.} 
  \label{msx_+_inc}
\end{figure*}

\begin{figure*}[t]
\center
  \includegraphics[scale=0.61]{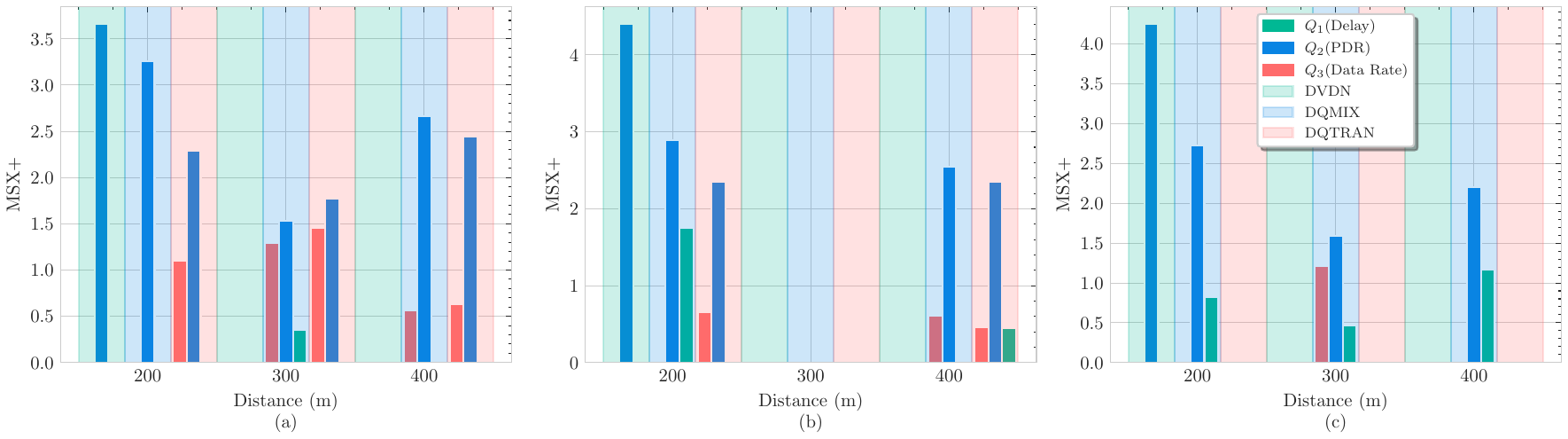} 
  \caption{Average MSX+$(``dec")$ for DQTRAN, DQMIX, and DVDN for three distances $200, 300, 400$ m: $\bm{(a)}$ AR agent, $\bm{(b)}$ VR agent, and $\bm{(c)}$ CG agent.} 
  \label{msx_+_dec}
\end{figure*}

In Fig. \ref{msx_+_inc}, we present the average positive minimal sufficient explanations ($MSX^+$) for the selected codec parameter compared to selecting a higher configuration value. It can be observed that for certain distances, only $Q_2$ (PDR) is necessary to explain the chosen action. Additionally, it is noticeable that for certain algorithms, $(MSX^+) = \{\}$, indicating that all components are equally important. In Fig. \ref{msx_+_dec}, we illustrate the ($MSX^+$) for the selected codec parameter over selecting a lower configuration value. This figure demonstrates a similar trend as Fig. \ref{msx_+_inc}, albeit with the distinction that a lower configuration parameter assigns greater importance to the DRR. In contrast, a higher codec configuration prioritizes delay. Intuitively, this result is expected, as reducing the codec parameter allows more flexibility to increase the data rate, whereas selecting a higher parameter emphasizes the delay component. The graph of the average negative minimal sufficient explanations $(MSX^-)$ is not included in this work, due to $(MSX^-) = \{\}$, which means that all of the agents believe that the action they have selected dominates over any fluctuation of the codec parameters. Note that, these results are solely related to this case study, but the method and framework proposed in this work can be employed in any other MARL algorithm. 

\subsection{Importance reward metric}
\begin{figure}[t]
\center
  \includegraphics[scale=0.75]{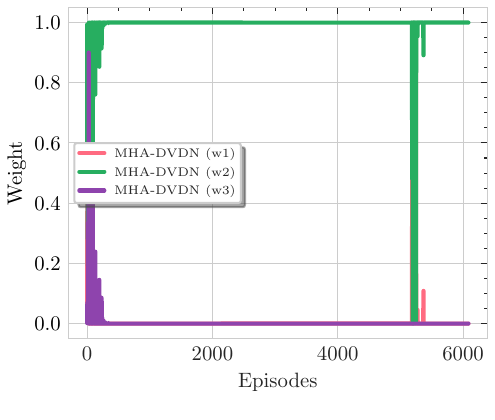}
  \setlength{\belowcaptionskip}{-5pt}
  \caption{Behavior of importance weights in Eq. (\ref{no_variance_weights}) when considering no variance. $w1, w2, w3$ corresponds to the weights of the reward components, delay, PDR, and DRR, respectively. The figure presents the behavior of the weights for the Multi-Headed Adaptive DVDN algorithm when users are located at a distance of $300$ m.} 
  \label{no_variance_weights_graph}
\end{figure}

\begin{figure}[t]
\center
  \includegraphics[scale=0.75]{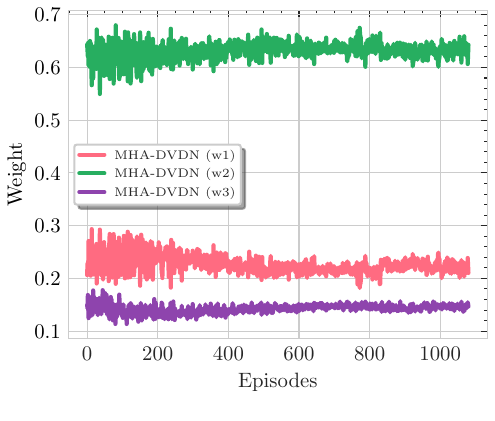}
  \setlength{\belowcaptionskip}{-5pt}
  \caption{Behavior of importance weights in Eq. (\ref{weight_reward_episode}) when considering variance. $w1, w2, w3$ corresponds to the weights of the reward components, delay, PDR, and DRR, respectively. The figure presents the behavior of the weights for the Multi-Headed Adaptive DVDN algorithm when users are located at a distance of $300$ m.} 
  \label{variance_weights_graph}
\end{figure}

In this subsection, we present the results that led us to the design of the importance reward metric utilized as weights in the adaptive decomposed algorithms. The weights computed per episode tend to have a preference for the best reward in terms of magnitude, as observed in Fig. \ref{no_variance_weights_graph}. In such a figure, we present as an example the behavior of the weights during a training session of MHA-DVDN\footnote{All adaptive algorithms (MHA-DQMIX and MHA-DQTRAN) showed a similar behavior.} at $300$ m. As observed, $w2$ corresponding to the reward component PDR is always given maximum importance with respect to the rest of the weights $w1$ and $w3$, delay, and DRR, respectively. This is undesirable, leaving no room for other components to contribute to the learning process and making training unstable. To solve this issue we include the variance term and use $f(x) = \sinh{x}; x > 0$ to add importance to those reward components with high variance throughout training. In Fig. \ref{variance_weights_graph}, it can be seen how such modification helps to give more importance to low mean value rewards. 

\subsection{Performance of the proposed decomposed MARL algorithms}
\begin{figure*}[t]
\center
  \includegraphics[scale=0.51]{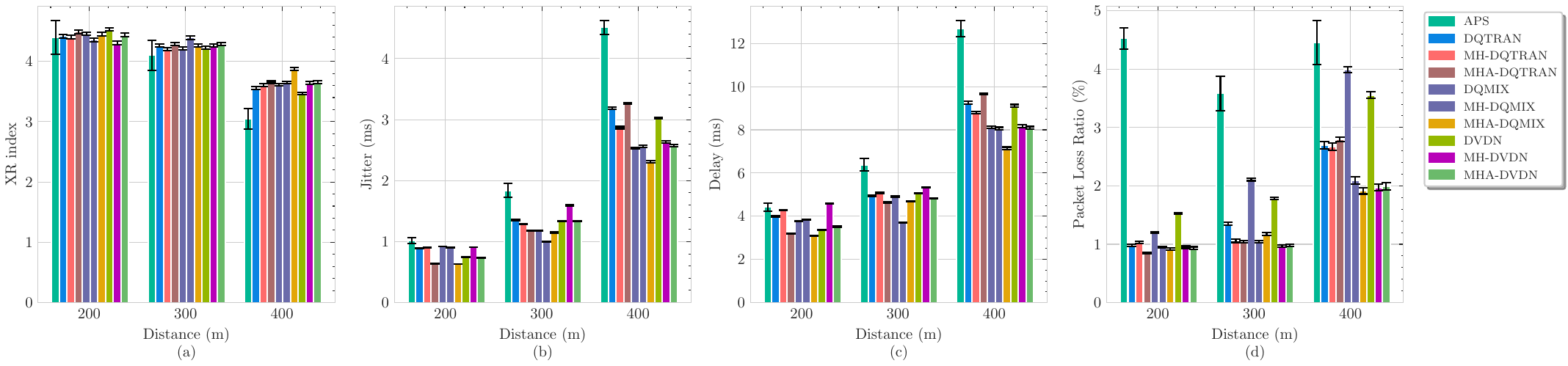} 
  \caption{Performance comparison between the baseline APS and the proposed decomposed reward MARL algorithms DQTRAN, DQMIX, and DVDN for three distances $200, 300, 400$ m: $\bm{(a)}$ XR index, $\bm{(b)}$ jitter, $\bm{(c)}$ delay, and $\bm{(d)}$ packet loss ratio.} 
  \label{APS_DVDN_DQMIX_DQTRAN}
\end{figure*}

In this subsection, we demonstrate that reward decomposition not only provides valuable insights into the design of MARL algorithms but also shows that the decomposed architecture can be utilized to enhance network performance while offering additional information. In Fig. \ref{APS_DVDN_DQMIX_DQTRAN}, we present comparative graphs illustrating performance in terms of network KPIs: XR index, jitter, delay, and PLR, respectively. Each subplot showcases the performance achieved by distance and the proposed VFF decomposed algorithms alongside the baseline APS. It can be observed that when UEs are closer to their respective base station ($d \in \{200,300\}$ m), the performance is similar, except in terms of PLR. However, the difference becomes significant as the distance increases to $400$ m, corresponding to scenarios where UEs are more impacted by propagation path loss. Additionally, we provide the average percentage gains over the APS algorithm in Table \ref{compartative_table}. Indicated by a green arrow up $\color{green}(\uparrow)$ and a blue arrow down $(\color{blue}\downarrow)$, we denote the percentage increase or decrease of the network KPIs. For example, we observe that in most cases, the variants corresponding to QMIX perform the best (indicated by bold font). Furthermore, architectural changes inspired by multi-task learning contribute positively to KPI improvements across almost all algorithms. The best algorithm, MHA-DQMIX, offers an average improvement of up to $10.7 \%$, $41.4 \%$, $33.3 \%$, and $67.9\%$ in terms of XR index, jitter, delay, and PLR, respectively.

\subsection{Potential applications in the telecommunications industry}
XRL-based reward decomposition techniques have great potential for telecommunications companies. Below, we list some potential future applications and direct applications in the industry:

\begin{itemize}
    \item \textbf{RL algorithm selection}: Given $M$ RL algorithms, a 6G/ML engineer can observe how reward design in the Markov Decision Process (MDP) affects each algorithm. This observation can lead to an informed selection of the best algorithm.
    \item \textbf{Post-hoc RL algorithm failure analysis}: At a failure event, reward decomposition allows post-hoc analysis of the performance of RL algorithms by inspecting the decomposed rewards in each observed state. Note that a logging system must be operating to enable such analysis.
    \item \textbf{Reward design in RL algorithms}: Reward decomposition allows the optimization of RL algorithms by analyzing the importance of each component of the composite reward objective function. This can potentially lead to a reduction in the reward composition size and, consequently, a decrease in the number of Application Programming Interface (API) requests to the cloud, vendor, or analytics provider to obtain any KPIs utilized to derive each reward.
\end{itemize}

\begin{table}
\caption{Comparison between the baseline APS and the proposed algorithms in the percentage (\%) of increase $\color{green}(\uparrow)$ and decrease $(\color{blue}\downarrow)$.}

\begin{center}
\resizebox{\columnwidth}{!}{%
\begin{tabular}{c c c c c } 
\hline
\textbf{Algorithm}&\textbf{XR Index}$\color{green}(\uparrow)$&\textbf{Jitter}$(\color{blue}\downarrow)$&\textbf{Delay}$(\color{blue}\downarrow)$&\textbf{PLR}$(\color{blue}\downarrow)$ \\
\hline
DQTRAN & {$7.0$} &$22.5$ & $19.7$ & $\bm{60.1}$\\
MH-DQTRAN & {$7.0$} & $25.8$ & $18.0$  & $62.6$\\
MHA-DQTRAN & {$8.8$} & $33.6$ & $26.3$ & $63.1$\\
\hline
DQMIX & {$\bm{7.5}$} & $\bm{29.8}$ & $\bm{24.6}$ & $41.7$\\
MH-DQMIX &  {$\bm{8.6}$} & $\bm{33.3}$ & $\bm{30.7}$  & $67.6$\\
MHA-DQMIX & {$\bm{10.7}$} & $\bm{41.4}$ & $\bm{33.3}$ & $67.9$\\
\hline
DVDN & {$6.5$} & $28.9$ & $24.3$ & $45.6$\\
MH-DVDN &  {$7.1$} & $21.8$ & $16.2$ & $\bm{69.2}$\\
MHA-DVDN & {$8.4$} & $32.9$ & $27.0$ & $\bm{69.1}$\\  
\hline
\end{tabular}
}
\label{compartative_table}

\end{center}
\vspace{-7mm}
\end{table}

\section{Conclusions } \label{Section6}
In this paper, we propose  Value Function Factorization (VFF)-based Explainable Multi-Agent Reinforcement Learning (XMARL) algorithms that utilize reward decomposition in XR codec adaptation problems. We present theoretical bounds that VFF-XMARL algorithms can be decomposed and we introduce architectural changes to several VFF-XMARL algorithms, such as Value Decomposition Networks (VDN), QMIX (Mixture of Q-Values), and QTRAN (Q-Value Factorization). Additionally, we propose utilizing ideas from multi-task learning to reduce the overhead of vanilla decomposed MARL algorithms. Moreover, we introduce a novel metric called Reward Difference Fluctuation Explanations (RDFX), which is suitable when actions correspond to configuration metrics typically found in wireless network problems. Furthermore, we present a solution that allows providing explanations in an online manner and using them to improve current algorithms, named Multi-Headed Adaptive. This proposal utilizes network gradients and leverages reward decomposition to enhance the action selection of the XMARL algorithms. We observe that for XR codec adaptation, the most crucial reward component corresponds to the packet delivery ratio (PDR), according to the minimal sufficient explanation (MSX) metric. In conclusion, the utilization of MARL algorithms using reward decomposition in wireless networks not only provides a better understanding of the design of composite rewards but also can improve trust and network performance if careful design is performed.

\section{Acknowledgment }\label{Section7}
This research is supported by the NSERC Canada Research Chairs program, Mitacs Accelerate Program, and Ericsson.\vspace{-2mm}



\bibliography{biblio.bib}{}
\bibliographystyle{IEEEtran}

\begin{appendices}
\section{Proofs}
\label{Section8}
The following proof shows how reward decomposition can be applied to value factorization multi-agent algorithms.
\subsection{Useful Definitions}
Let a multi-agent Markov Game (MAMG) take the form of the following tuple $\mathcal{M}\mathcal{G}(\mathcal{S}, \mathcal{A}, P, r, N, \gamma, d_0)$ where $\mathcal{S}$ corresponds to the joint state space, $\mathcal{A}$ the joint action space, $P$ the transition probability function with $P(\cdot|\bm{s}, \bm{a})$ giving the distribution over states if the joint action space at time $t$, $\bm{a} = <a_1, ..., a_N >$ is taken at the joint state $\bm{s} = <s_1, ..., s_N>$. $r$ corresponds to the team reward $r:\mathcal{S} \times \mathcal{A}$. $\gamma$ is the discount factor. $N$ is the number of agents and $d_0$ corresponds to the initial state distribution. At each time step, all agents take individual actions according to their policy $\pi_i(\cdot_i, s_t)$. This gives a joint action policy $\bm{\pi}(\cdot|\bm{s}_t)\stackrel{\text{def}}{=} \prod_{i\in N}\pi_i(\cdot_i|s_i)$ with a joint action $\bm{a}_t \in \mathcal{A}$ at state $\bm{s}_t \in \mathcal{S}$. The joint state-action value function $\bm{Q}^{\pi}: \mathcal{S} \times \mathcal{A} \rightarrow r \stackrel{\text{def}}{=} [Q_{min}, Q_{max}]$ can be defined as, 

\begin{equation}
    \bm{Q}^{\bm{\pi}} \stackrel{\text{def}}{=} \mathbb{E}_{s_{1:\infty} \sim P, \bm{a}_{1:\infty} \sim \bm{\pi}}\left[\sum_{t=0}^\infty \gamma^t r(\bm{s}_t, \bm{a}_t)| \bm{s}_0 = \bm{s},\bm{a}_0 = \bm{a} \right]
\end{equation}

\textbf{Definition A.1}: \textit{A multi-agent Markov Game $\mathcal{M}\mathcal{G}(\mathcal{S}, \mathcal{A}, P, r, N, \gamma, d_0)$ is considered a decomposable game if its team objective function $r: \mathcal{S} \times \mathcal{A}$ takes the form, }
\begin{equation} \label{dec_rewards}
    r(\bm{s}, \bm{a}) = r_1(s_1, a_1) + ...+ r_N(s_N, a_N),  
\end{equation}
\textit{where $r_i: \mathcal{S}_i \times \mathcal{A}_i$ is the independent reward for the $i^{th}$ agent. This definition allows the decomposition of the transition function $P$ as,}
\begin{equation}\label{dec_probs}
    P(\bm{s}'| \bm{s}, \bm{a}) = F_1(\bm{s}'| s_1, a_1) + ...+ F_N(\bm{s}'| s_N, a_N),  
\end{equation}
Consequently, substituting Equations \eqref{dec_rewards} and \eqref{dec_probs} in the Bellman equation, 
\begin{align}
    \bm{Q}^{\bm{\pi}} &= r(\bm{s}, \bm{a}) + \gamma\mathbb{E}_{\bm{s}'\sim P}[V^{\bm{\pi}}(\bm{s}')]\\\notag
     &= r(\bm{s}, \bm{a}) + \gamma \int_{\mathcal{S}}V^{\bm{\pi}}(\bm{s}')P(\bm{s}'|\bm{s}, \bm{a})d\bm{s}'\\ \notag
     &= \sum_i^Nr_i(s_i,a_i) + \gamma \int_{\mathcal{S}}V^{\bm{\pi}}(\bm{s}')\sum_i^N F_i(\bm{s}'|s_i, a_i)d\bm{s}'\\\notag
     &= \sum_i^N \left[r_i(s_i,a_i) + \gamma \int_{\mathcal{S}}V^{\bm{\pi}}(\bm{s}') F_i(\bm{s}'|s_i, a_i)d\bm{s}'\right]\\   
     &\stackrel{\text{def}}{=} \sum_i^N Q_i^{\pi}(s_i, a_i) \label{dec_game_final} 
\end{align}
\subsection{Proof of Theorem \ref{theorem}}\label{appendix-proof}

 If the composite reward of agent $i^{th}$ is decomposed in $c^{th}$ components as follows, 
\begin{equation} \label{eq1_proof}
   r_i(s_{i}, a_{i}) = \sum_c^C \sigma_{c,i}r_{c,i}(s_{i},a_{i}).
\end{equation}
The Q-value function for the $i^{th}$ agent and $c^{th}$ reward component can be defined as, 
\begin{equation} \label{eq1_proof1}
    Q_{c,i}^{\pi} \stackrel{\text{def}}{=}  \mathbb{E}_{s_{1:\infty} \sim P_i, a_{1:\infty} \sim \pi_i}\left[\sum_{t=0}^\infty \gamma^t r_{c,i}(s_{t,i}, a_{t,i})| s_{0,i} = s_i, a_{0,i} = a_i \right].
\end{equation}

The Q-value function for the $i^{th}$ agent can be defined as, 
\begin{equation} \label{eq2_proof}
    Q_{i}^{\pi} \stackrel{\text{def}}{=}  \mathbb{E}_{s_{1:\infty} \sim P_i, a_{1:\infty} \sim \pi_i}\left[\sum_{t=0}^\infty \gamma^t r_{i}(s_{t,i}, a_{t,i})| s_{0,i} = s_i, a_{0,i} = a_i \right].
\end{equation}
For the sake of abbreviation, we use term $\mathbb{E}_{{\pi}_i}$  instead of $\mathbb{E}_{s_{1:\infty} \sim P_i, a_{1:\infty} \sim \pi_i}$. Substituting \eqref{eq1_proof} in \eqref{eq2_proof} we have, 
\begin{align}
     Q_i^{\pi} &=  \mathbb{E}_{{\pi}_i}\left[\sum_{t=0}^\infty \gamma^t \sum_c^C \sigma_{c,i}r_{c,i}(s_{t,i}, a_{t,i})\right] \\\notag
     &= \mathbb{E}_{{\pi}_i}\left[\sum_{t=0}^\infty \sum_c^C \gamma^t  \sigma_{c,i}r_{c,i}(s_{t,i}, a_{t,i})\right] \\\notag
     &= \mathbb{E}_{{\pi}_i}\left[\sum_c^C \sum_{t=0}^\infty  \gamma^t  \sigma_{c,i}r_{c,i}(s_{t,i}, a_{t,i})\right] \\\notag
     &=  \mathbb{E}_{{\pi}_i}\left[\sum_c^C \sigma_{c,i} \sum_{t=0}^\infty  \gamma^t  r_{c,i}(s_{t,i}, a_{t,i})\right] \\\notag 
     &= \sum_c^C  \sigma_{c,i}\mathbb{E}_{{\pi}_i}\left[ \sum_{t=0}^\infty  \gamma^t  r_{c,i}(s_{t,i}, a_{t,i})\right] \\ 
     &= \sum_c^C  \sigma_{c,i}Q_{c,i}^{\pi} \label{eq3_proof}
\end{align}
Finally, substituting \eqref{eq3_proof} in \eqref{dec_game_final} gives, 
\begin{align}
    \bm{Q}^{\bm{\pi}}(s, \bm{a})  =  \sum_i^N  \sum_c^C  \sigma_{c,i}Q_{c,i}^{\pi}(s, \bm{a}), \label{proof_final} 
\end{align} which finalizes the proof. 

\end{appendices}
\end{document}